\documentclass[fleqn,usenatbib]{mnras}

\usepackage{newtxtext,newtxmath}
\usepackage[T1]{fontenc}

\DeclareRobustCommand{\VAN}[3]{#2}
\let\VANthebibliography\thebibliography
\def\thebibliography{\DeclareRobustCommand{\VAN}[3]{##3}\VANthebibliography}

\usepackage{graphicx}	% Including figure files	% Advanced maths commands

\usepackage{amsmath}

\usepackage{bm}
\usepackage{float}
\usepackage{array}
\usepackage{makecell}
\usepackage{multirow}
\usepackage{comment}
\usepackage{url}      % for \url in footnotes
\usepackage{hyperref} % load near end of packages; adjust colors if desired

\newcommand{\HI}{\rm H{\sc i}}
\newcommand{\HII}{\rm H{\sc ii}}

\newcommand{\XHI}{$x_{\rm HI}$}
\newcommand{\AAstar}{AA$^{\star}$} 

\title[Ionized bubble detection]{Probing ionized bubbles around luminous sources during  reionization with SKA 21-cm observations}

\author[A. Mishra et al.]{
Arnab Mishra$^{1}$\thanks{E-mail: arnabm.physics.rs@jadavpuruniversity.in},
Kanan K. Datta$^{1}$\thanks{E-mail: kanankdatta.physics@jadavpuruniversity.in},
Chandra Shekhar Murmu$^{2,3,4}$,
Samir Choudhuri$^{5}$,
Iffat Nasreen$^{1}$, and
\newauthor 
Snehasish Saha$^{1}$\\
$^{1}$Relativity and Cosmology Research Centre (RCRC), Department of Physics, Jadavpur University, Kolkata 700032, India\\
$^{2}$Department of Astronomy, Astrophysics and Space Engineering, Indian Institute of Technology Indore, Khandwa Road, Simrol, Indore 453552, India\\
$^{3}$Astrophysics Research Center of the Open University (ARCO), The Open University of Israel, P.O. Box 808, 1 University Road, Ra’anana 4353701, Israel.\\
$^{4}$Department of Natural Sciences, The Open University of Israel, P.O. Box 808, 1 University Road, Ra’anana 4353701, Israel.\\
$^{5}$Centre for Strings, Gravitation and Cosmology, Department of Physics, Indian Institute of Technology Madras, Chennai 600036, India
}

\date{Accepted XXX. Received YYY; in original form ZZZ}

\pubyear{\the\year{}}

\begin{document}
\label{firstpage}
\pagerange{\pageref{firstpage}--\pageref{lastpage}}
\maketitle

% Abstract of the paper
\begin{abstract}
Detecting and characterizing individual ionized bubbles during the Epoch of Reionization (EoR) using the redshifted \HI\ 21-cm signal provides a direct probe of the early ionizing sources and the intergalactic medium. We develop and validate a computationally efficient estimator that operates on gridded visibilities to detect ionized bubbles. This serves as an accurate alternative to the more computationally demanding bare estimator that uses all baselines and frequency channels. Further, we employ a non-parametric foreground-subtraction method based on Gaussian process regression, which minimizes loss of the \HI\ 21-cm signal and yields improved signal-to-noise ratios. Our analysis indicates that ionized bubbles at redshifts $z \sim 7 - 8$ can be detected with SNR $\gtrsim 10$ using $\sim 100$ hours of SKA1-Low \AAstar\ and AA4 observations. We further derive a scaling relation that connects the SNR to the bubble radius, redshift, total observing time, and the mean neutral hydrogen fraction of the surrounding IGM. This helps to quickly predict the observational outcome for any planned observations and is, therefore, useful for devising observational strategies. Finally, we apply a Bayesian likelihood framework with Markov Chain Monte Carlo sampling to the residual visibilities to recover ionized bubble properties, including radius, position, and the mean neutral fraction. The resulting posterior distributions demonstrate accurate recovery of the bubble parameters. This confirms the feasibility of robustly characterizing individual ionized regions with the SKA1-Low.

\end{abstract}

% Select between one and six entries from the list of approved keywords.
% Don't make up new ones.

%%%%%%%%%%%%%%%%%%%%%%%%%%%%%%%%%%%%%%%%%%%%%%%%%%

%%%%%%%%%%%%%%%%% BODY OF PAPER %%%%%%%%%%%%%%%%%%

\section{Introduction}
\label{sec:intro}

The first galaxies and quasars began to ionize the neutral hydrogen (\HI) in the intergalactic medium (IGM) during the epoch of reionization, which occurred over the redshift range $z \sim 15$ to $6$ \citep{Pritchard_2012, bera2023}. This process led to the formation of fully ionized regions, commonly referred to as ionized bubbles, around early luminous sources. As reionization progressed, these bubbles grew in size and eventually merged, completing the ionization of the IGM \citep{mellema2006, Furlanetto_2006, tirth2009}.\par

Recent surveys have reported detections of bright quasars during the EoR \citep{Mortlock_2011, Wu_2015, Ba_ados_2017, Wang_2018, Wang_2019, Matsuoka_2019a, Yang_2020, Wang_2021} and luminous galaxies at high redshifts \citep{Matsuoka_2019, witstok24}. Such sources are expected to produce large \HII\ regions in the surrounding IGM.  At the initial and intermediate stages of reionization, isolated or nearly isolated ionized bubbles are expected, which will remain buried inside the neutral (or partially ionized) IGM. Observations of these ionized bubbles in redshifted \HI\ 21-cm maps provide a direct probe of the epoch of reionization and offer valuable insights into the nature of the underlying ionizing sources. The study of individual ionized bubbles complements the widely explored approach of probing reionization through the power spectrum and other statistical measures \citep{Majumdar_2011, zack2020}. However, detecting this faint cosmological signal is challenging due to the overwhelming brightness of astrophysical foregrounds \citep{Di_Matteo+2002, Oh_2003, Santos+2005, Ali_2008} and instrumental noise, which are several orders of magnitude stronger than the 21-cm signal. The requirements for high-precision calibration and effective radio-frequency interference (RFI) mitigation further complicate these challenges \citep{samitpal2024, saikat2025}. \par

Prospects of detecting ionized bubbles in HI 21-cm images have been explored in several studies \citep{geil2008, mellema2013, Kakiichi_2017, giri2018, giri2018a, bianco2024, bianco2024a}. However, direct imaging of the IGM during the EoR via \HI\ 21-cm signal remains challenging for current facilities such as the uGMRT, LOFAR, and MWA due to their limited sensitivities.  While SKA1-Low is expected to image individual ionized bubbles, achieving this will require substantially long integration times \citep{Koopmans_2015, 10.1093/mnras/stw2494}.\par

Matched-filter techniques are often used to detect a weak signal of known functional form buried in strong noise. This technique combines the signal optimally using an appropriate filter, maximizes the signal-to-noise ratio, and thus reduces the total observation time considerably. This technique has been successfully used to detect gravitational waves \citep{Abbott2016}. %effective for detecting and characterizing individual ionized bubbles around known sources \citep{Datta_2007, Datta_2008, majumdar2012}, particularly when the system noise dominates over the target \HI\ signal and the signal’s functional form is known. 
The idea of applying a matched filter technique to detect individual ionized bubbles using HI 21-cm signal was first proposed in \cite{Datta_2007}. It presents a visibility-based framework to study the feasibility of detecting individual ionized bubbles through radio-interferometric observations of redshifted \HI\ 21-cm radiation. Later, it was shown that fluctuations in the HI density in the IGM outside the targeted ionized bubble behaves as noise and hinders the detection of ionized bubbles of small sizes with radius $\lesssim 6$ Mpc \citep{Datta_2008}. Subsequent work, such as \cite{Datta_2012, majumdar2012}, carried out a detailed study using numerical simulations. A scaling relation, which enables us to quickly estimate detection prospects for various ionized bubble sizes, redshifts, and instruments, was presented in \cite{Datta_2009}. Further, \cite{Ghara_2020} explored the possibility of constraining the parameters that characterize the bubble and IGM  using a Bayesian analysis. In our earlier work \citep{Mishra:2024jjg}, we studied the impact of foreground subtraction on the detectability of ionized bubbles using the matched-filter method with the uGMRT and SKA1-Low observations.  That study explicitly included simulations of the cosmological \HI\ 21-cm signal, realistic foreground contaminants, and system noise, followed by a foreground subtraction stage, making the detection analysis more representative of real observational scenarios. 

In this work, we present and validate a computationally efficient fast estimator that operates on gridded visibilities. This estimator is useful for analyzing large datasets from SKA1-Low. We simulate realistic mock datasets specific to SKA1-Low array layouts (AA2, AA*, and AA4) that include the cosmological \HI\ signal, astrophysical foregrounds, and system noise. We then employ a new non-parametric method for foreground subtraction based on Gaussian regression process (GPR). We have shown this method to be significantly more effective than traditional polynomial-based methods. The GPR is extensively used in Machine learning and has been used in the EoR study to successfully subtract foreground and other systematics \citep{mertens2018,Ghosh2020}. We also present a scaling relation, specific to the SKA1 -Low, between the signal-to-noise ratio (SNR) and other parameters, such as bubble radius, redshift, observation time, and mean neutral hydrogen fraction in the IGM. Finally, we perform Bayesian inference using Markov Chain Monte Carlo (MCMC) techniques \citep{Ghara_2020} to recover bubble parameters, including radius, spatial and line of sight positions,  and the surrounding neutral fraction.  
\par
Throughout our analysis we use the cosmological parameters $ h= 0.7$, $\Omega_{\rm m} = 0.27$, $\Omega_{\Lambda} = 0.73$, $\Omega_{\rm b} = 0.044$, ${\sigma_8=0.83}$~ and ${n_{\rm s}=0.96}$ consistent with the WMAP measurements~\citep{Bennett_2013}.

\section{Simulating mock observations}
\label{sec:simulation}

Here, we briefly describe our simulations used to generate mock data that resemble SKA1-Low observations at frequencies relevant to the EoR. These simulations include the generation of the \HI\ 21-cm maps around bright quasars/galaxies, along with astrophysical foregrounds. Subsequently, these simulated maps are converted into visibilities, and the system noise is added to mimic SKA1-Low-like observations. Below, we describe them in detail.

\subsection{Observation setup}
\label{subsec:obs_setup}

Our simulations focus on two observing frequencies, $175$~MHz and $153$~MHz, corresponding to redshifts of $z \approx 7.1$ and $z \approx 8.3$ for \HI\ 21-cm line observations. The selection of these two frequencies is motivated by the discoveries of a bright QSO at redshift $z=7.1$~\citep{Mortlock_2011} and a large ionized bubble at redshift~$z=8.3$~\citep{witstok24}. The angular size of our simulation cube is set to $\sim 3.3^\circ \times 3.3^\circ$, which is closer to the SKA1-Low primary field of view at the observing frequencies considered here. Our simulation box has a total of $512$ grids on each side with a resolution of $1.03$ Mpc comoving length, which results in $527$ Mpc on each side. This results in an angular resolution of $\sim 23^{\prime\prime}$. In our analysis, we have used $256$ frequency channels with a total bandwidth of $16$ MHz. This results in a frequency channel width of $\Delta \nu= 62.5$ kHz.

\begin{figure*}
\centering
\includegraphics[width=0.33\textwidth]{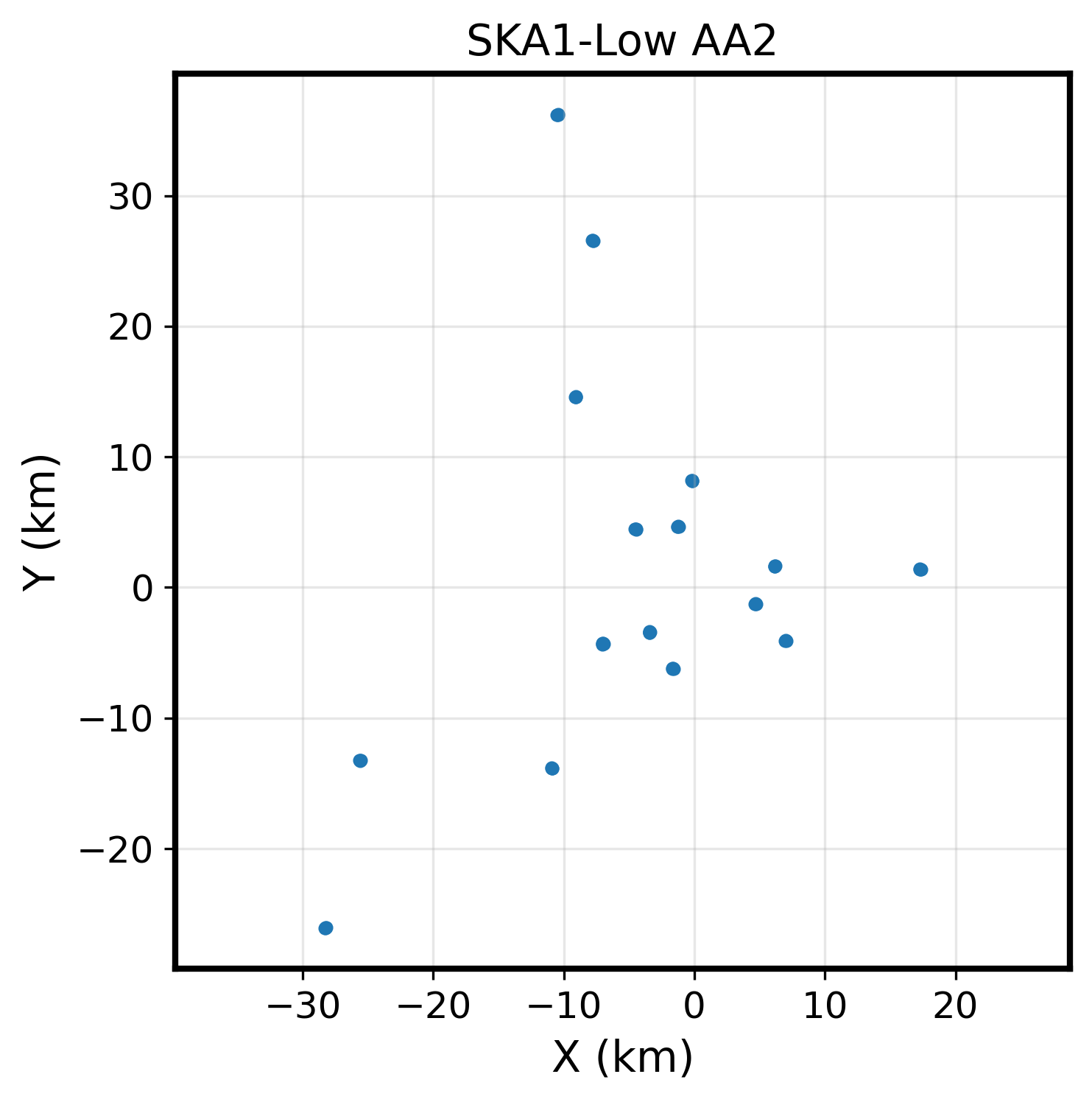}
\includegraphics[width=0.33\textwidth]{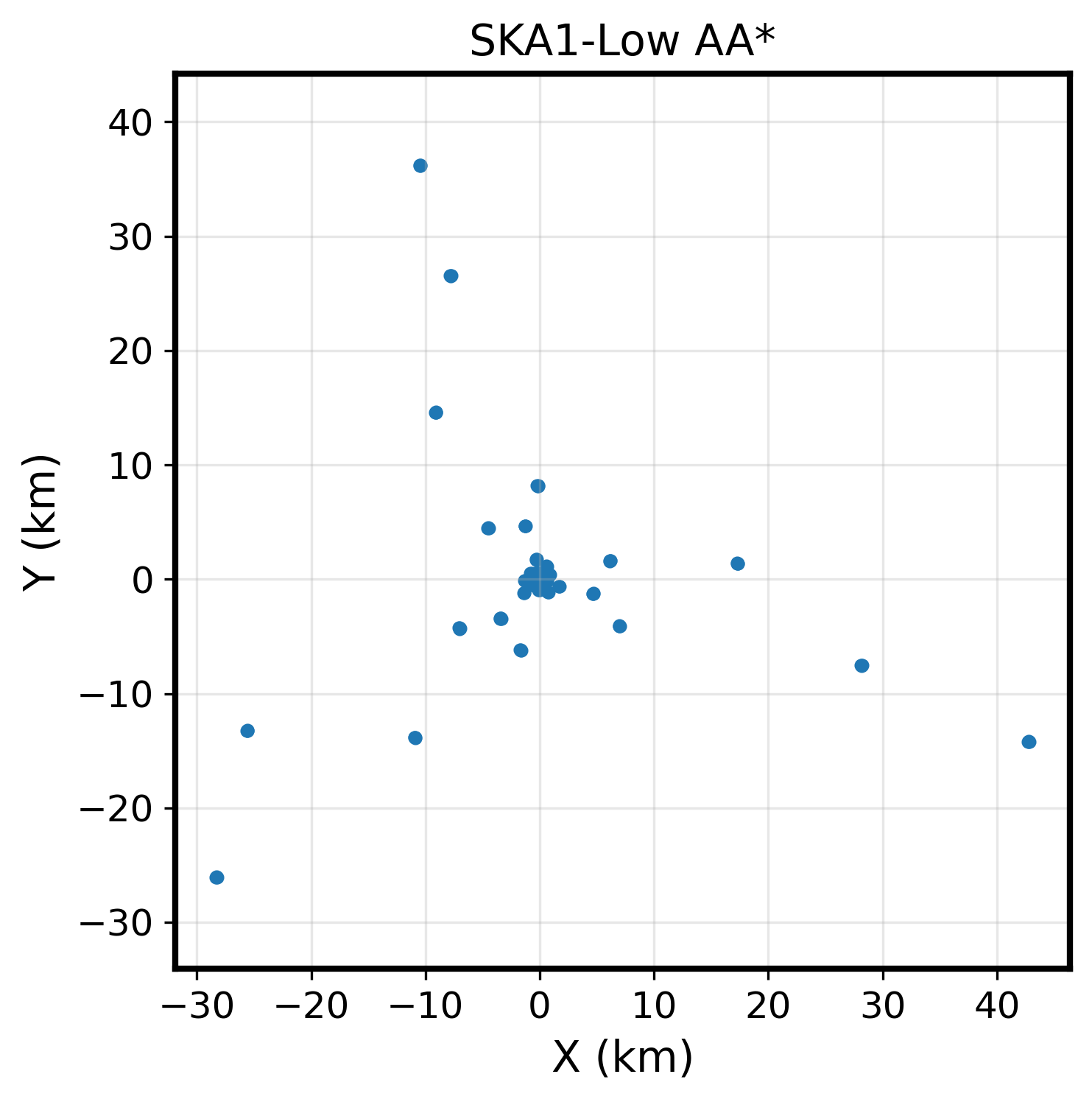}
\includegraphics[width=0.33\textwidth]{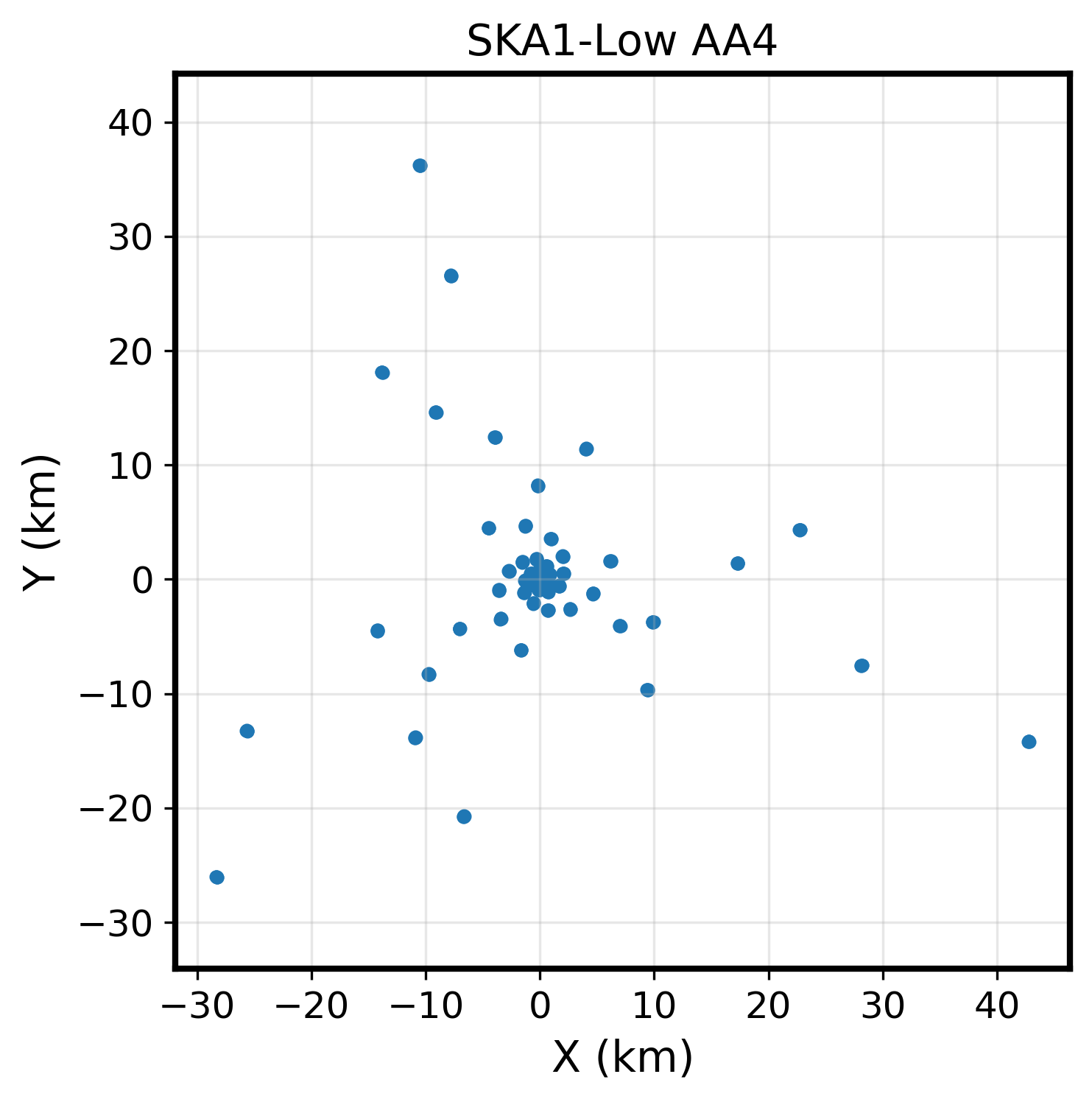}
\caption{Station layouts of the SKA1-Low configurations: AA2 (left), \AAstar\ (middle), and AA4 (right).}
\label{fig:ska_array_config}
\end{figure*}

\begin{figure*}
\centering
\includegraphics[width=0.33\textwidth]{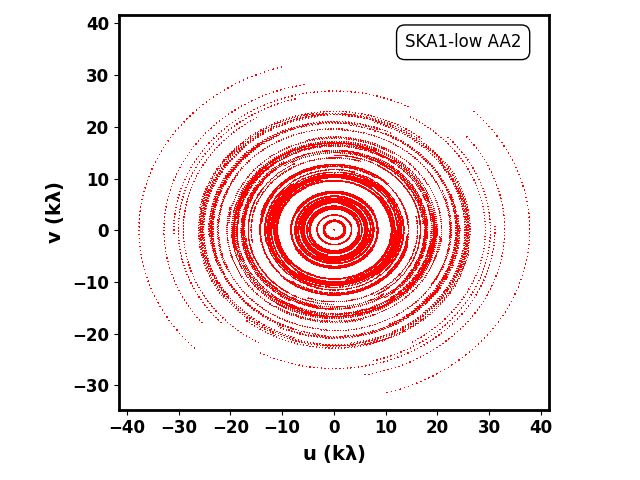}
\includegraphics[width=0.33\textwidth]{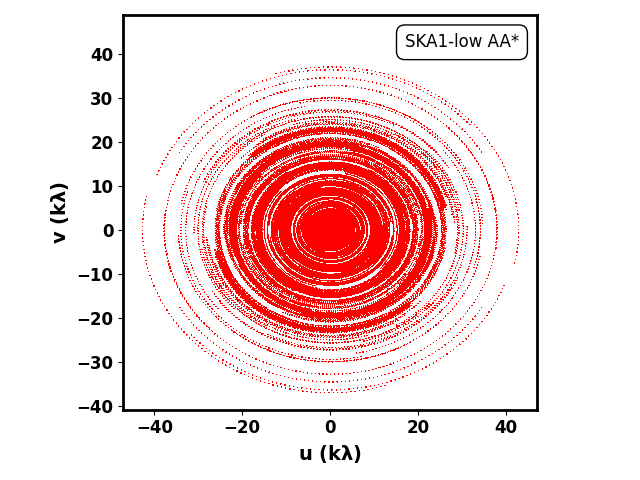}
\includegraphics[width=0.33\textwidth]{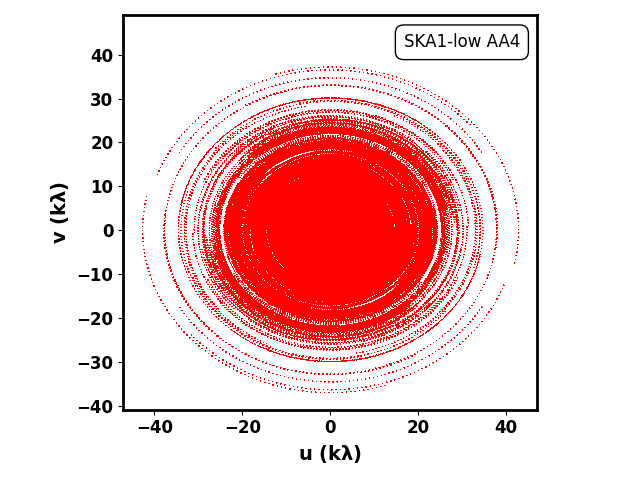}
\caption{$uv$-coverage for $8$ hours of observation at $\nu = 175$ MHz for the SKA1-Low configurations AA2 (left), \AAstar\ (middle), and AA4 (right). }
\label{fig:Baseline_distribution}
\end{figure*}

We consider three SKA1-Low configurations, representing successive stages of its phased deployment: AA2, \AAstar\, and AA4. These stages will include $68$, $307$, and $512$ stations, respectively \citep{braun2019anticipatedperformancesquarekilometre}. Figure~\ref{fig:ska_array_config} shows the station layouts. The AA2 represents the early deployment with sparse coverage, especially in the core. In contrast, the \AAstar\ has much denser concentration of stations in the core, while the full AA4 configuration extends the array with spiral arms. This provides both dense core coverage and long baselines. 
The baseline distributions for these configurations for $8$ hours of observation at $\nu=175$~MHz, assuming a correlator integration time of $320$ seconds and a declination of $+06^\circ$, are shown in Figure~\ref{fig:Baseline_distribution}. AA2 provides relatively poor coverage at both short and long baselines due to its small number of stations. \AAstar\ fills the central region more effectively due to its dense core, and AA4 delivers the best overall $uv$-coverage with both compact and extended baselines. 

\begin{figure}
\centering
\includegraphics[width=0.49\textwidth]{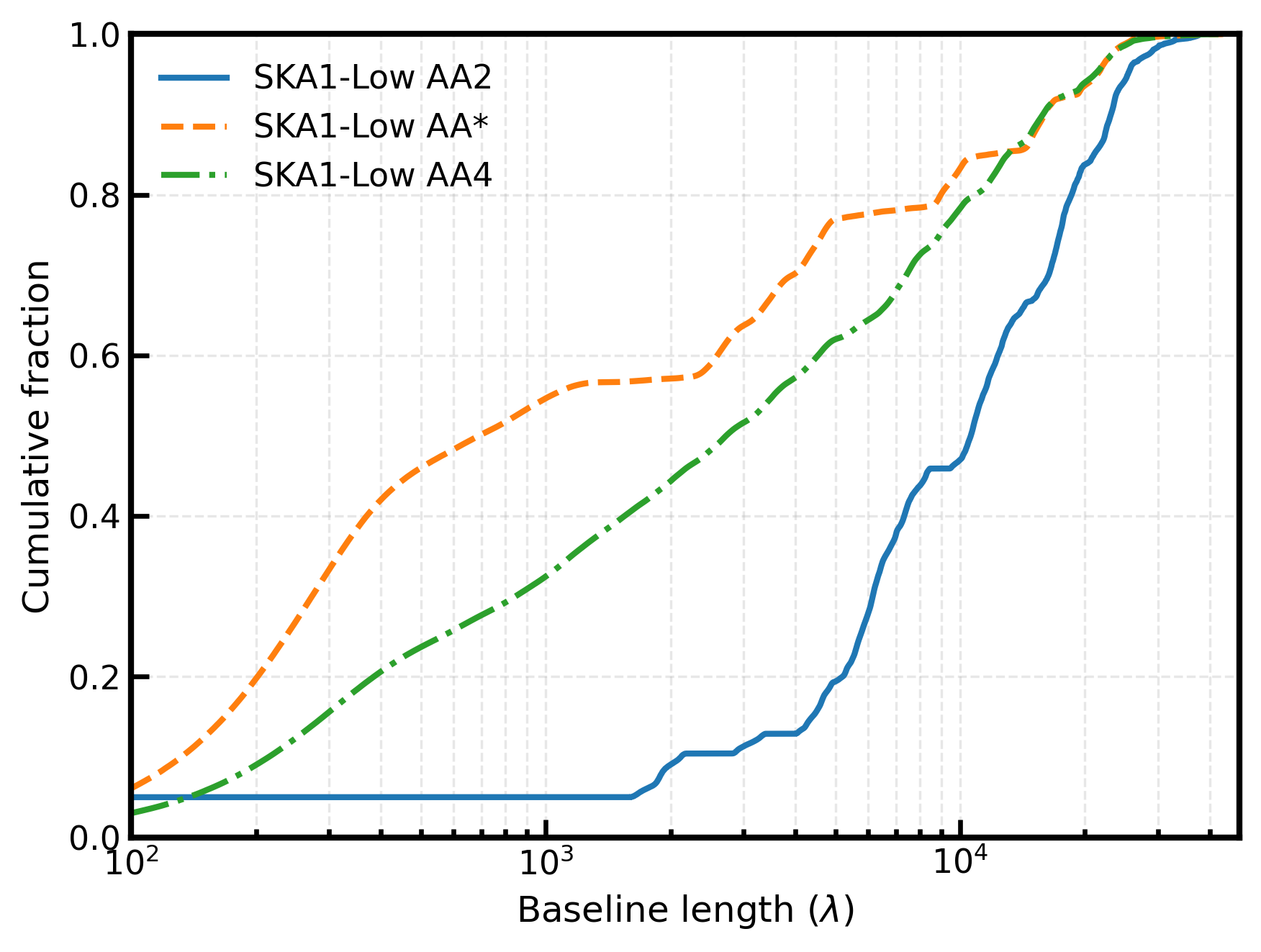}
\caption{Cumulative distribution function (CDF) of baseline lengths for the SKA1-Low configurations AA2, \AAstar\, and AA4 at $\nu=175$ MHz.}
\label{fig:cumulative_distribution}
\end{figure}

Figure~\ref{fig:cumulative_distribution} compares the distribution of baselines in all the SKA1-Low configurations across baseline lengths using the cumulative distribution function (CDF). The CDF for \AAstar\ rises most steeply, which confirms that its layout is dominated by short baselines. Approximately $60\%$ of its baselines are shorter than $2000 \,\lambda$. In contrast, AA2 has a much flatter distribution at shorter baselines, indicating a less pronounced concentration of short baselines (around $10\%$ of its baselines are shorter than $2000 \,\lambda$). The final AA4 configuration balances between shorter and longer baselines, creating a distribution intermediate between AA2 and \AAstar. This design ensures that AA4 will be sensitive to a wide range of angular scales.
Table~\ref{tab:sensitivity} lists the expected sensitivity parameter ($A_{\rm eff}/T_{\rm sys}$) for the SKA1-Low configurations at frequencies $\nu=150$ MHz and $170$ MHz, as provided in the latest SKA documentation \citep{braun2019anticipatedperformancesquarekilometre}. 

\begin{table}
\centering

\caption{Sensitivity parameters ($A_{\rm eff}/T_{\rm sys}$ in m$^2$ K$^{-1}$) for the SKA1-Low station and the three array configurations at the two observing frequencies. Values are taken from the SKA documentation \citep{braun2019anticipatedperformancesquarekilometre}.}
\label{tab:sensitivity}
\begin{tabular}{c c c c c}
\hline
\multirow{2}{*}{Frequency (MHz)} & \multirow{2}{*}{Single station} & \multicolumn{3}{c}{Full array configuration} \\
\cline{3-5}
 & & AA2 & \AAstar\ & AA4 \\
\hline
150 & 1.119 & 76.09 &343.5  & 572.8 \\
170 & 1.171& 79.63 & 359.5 & 599.7 \\
\hline
\end{tabular}

\end{table}

\subsection{HI 21-cm signal}
\label{subsec: HI_sim}
The \HI\ 21-cm signal around bright quasar/galaxy cluster is simulated following the same methodology described in our earlier work \cite{Mishra:2024jjg}, but adapted here to the different SKA1-Low observational setups. We generate dark matter distributions using the $N$-body code\footnote{\url{https://github.com/rajeshmondal18/N-body}} developed by \cite{Bharadwaj_2004a} and identify halos with a Friends-of-Friends~\footnote{\url{https://github.com/rajeshmondal18/FoF-Halo-finder}} (FoF)~\citep{Mondal_2015} algorithm. We produce ionization maps and corresponding HI differential brightness temperature cubes using a semi-numerical prescription \citep{Choudhury_2009, Majumdar_2014, Mondal_2017}. 

We consider the following three representative scenarios. The first corresponds to a spherical ionized bubble around a bright quasar at $z=7.1$ (observing frequency $175$ MHz). This is motivated by the discovery of a bright quasar reported in \cite{Mortlock_2011}.  The second scenario models a spherical ionized bubble around a galaxy cluster at $z=8.3$ (observing frequency $153$ MHz), as reported in \cite{witstok24}. The mean mass-averaged neutral hydrogen fractions in the first and second scenarios are $0.88$ and $0.94$, respectively. The third scenario is similar to the first but with a more patchy ionization environment and a mass-averaged neutral hydrogen fraction of $0.52$. The respective HI 21-cm images for all three scenarios can be found in Figure~1 of \citep{Mishra:2024jjg}. 

\subsection{Foregrounds}
\label{subsec:Foregrounds}
In addition to the cosmological \HI\ 21-cm signal, our simulation also includes the dominant astrophysical foregrounds, which are several orders of magnitude brighter than the former. As in our earlier study \citep{Mishra:2024jjg}, we account for two major foreground components in our simulation, i.e., diffuse galactic synchrotron emission (DGSE) and extragalactic point sources.
The DGSE is simulated as a statistically isotropic Gaussian random field. We model its angular power spectrum, $C^M_l(\nu)$ as a power-law as follows~\citep{Choudhuri_2014}:
\begin{equation}
C^M_l(\nu) = A_{150}\times {\left(\frac{1000}{l}\right)}^{\beta} {\left(\frac{\nu_0}{\nu}\right)^{2\alpha}},
\label{eq:dgse_angular_power_spectrum}
\end{equation}
where $l$ is the angular multipole, and the other parameters have the following values: $\beta=2.34$, $\alpha=2.8$ and $A_{150} = 513 \, \text{mK}^2$~\citep{10.1111/j.1365-2966.2012.21889.x, Santos+2005, Ali_2008}. 

The population of extragalactic point sources is generated based on the measured differential source count model at $150$ MHz~\citep{10.1111/j.1365-2966.2012.21889.x} and given by, 
\begin{equation}
    \frac{dN}{dS} = \frac{10^{3.75}}{\text{Jy\,sr}}\left(\frac{S}{1\,\text{Jy}}\right)^{-1.6}.
    \label{eq:diff_source_count}
\end{equation}
We consider flux densities between $0.1 - 1000$ mJy and assign spectral indices randomly between $0.7$ and $0.8$. In a $3.3^{\circ}\times 3.3^{\circ}$ field, this yields $7775$ point sources. Details can be found in \cite{Mishra:2024jjg}. 

We combine the simulated \HI\ 21-cm signal and the foreground maps to generate realistic sky maps across the observing bandwidth. Subsequently, these maps are Fourier-transformed to generate visibilities, which are then sampled at baselines corresponding to different SKA1-Low configurations. %(AA2, \AAstar, and AA4).

\subsection{Noise contribution}
To make our study more realistic, we introduce system noise contribution from radio interferometers in our simulated mock visibility data. The system noise contribution to the measured visibility in each baseline and frequency channel is an independent Gaussian random variable with zero mean and the root mean square \citep{Datta_2007} is as follows,

\begin{equation}
\sqrt{\langle N^2 \rangle} = \frac{\sqrt{2}k_BT_{\rm sys}}{A_{\rm eff}\sqrt{\Delta\nu\Delta t}},
\label{eq:sys-noise}
\end{equation}
where $k_B$ is the Boltzmann constant, $T_{\rm sys}/A_{\rm eff}$ is the system-equivalent flux density (SEFD), $\Delta\nu$ is the channel width, and $\Delta t$ is the correlator integration time. $A_{\rm eff}/T_{\rm sys}$ values used in our study are given in Table~\ref{tab:sensitivity} which are consistent with  a single SKA1-Low station~\citep{braun2019anticipatedperformancesquarekilometre}. For our chosen channel width of $\Delta\nu = 62.5$ kHz and an integration time of $\Delta t = 320$ sec, the rms noise in visibility is: $0.35$ Jy at $153$ MHz and $0.37$ Jy at $175$ MHz. In this study, we simulate the SKA1-Low baseline (uv) coverage for $8$ hrs observations with a correlator integration time of $320$ sec. However, longer observation time will be required to achieve a significant detection of the \HI\ 21-cm signal. Therefore, we assume that multiple similar observing nights will be required. In practice, we reduce the system noise rms by a factor $\sqrt{t_{\rm obs}/8\, {\rm hrs}}$ to predict our results for the total observation time of $t_{\rm obs}$. %We note that the rms noise will be reduced by a factor of $\sqrt{t_{\rm obs}/8\, \text{hrs}}$ \KKD:  {\bf should not it be 8 hrs instead of $\Delta t$ because you already simulated 8 hrs of observations } in where $t_{\rm obs}$ is the total observation time. 
The resulting rms noise in the image plane can be estimated as, 
\begin{equation}
\sigma_{\rm rms} \approx \frac{\sqrt{2}k_B T_{\rm{sys}}}{A_{\rm{eff}}\sqrt{B_\nu t_{\rm{obs}}N_{\rm{ant}}(N_{\rm{ant}}-1)/2}},
\label{eqn:image_noise}
\end{equation}
where $B_\nu$ is the total observing bandwidth, and $N_{\rm{ant}}$ is the total number of antennae in the array. %We consider different total observation times for each SKA1-Low configuration to explore a range of sensitivities.  We consider longer integration times for the less sensitive arrays to achieve more comparable results. The specific total observing times are: $1000$ hours for AA2, $200$ hours for \AAstar, and $100$ hours for AA4 array. 
%The simulated noise is then added to the combined \HI\ 21-cm signal and foreground visibilities.

\section{Foreground subtraction: Gaussian Process Regression}
\label{sec:FG_subtract}

In our previous study \cite{Mishra:2024jjg}, we employed a polynomial fitting technique to remove the smooth foreground component directly from the total visibilities. Although this approach successfully subtracted foreground contamination, we found that it also led to partial subtraction of the \HI\ 21-cm signal itself. This occurs because the higher-order polynomial function not only fits the spectrally smooth foregrounds but also fits a part of the \HI\ 21-cm signal. This leads to a considerable reduction in the signal-to-noise ratio. 
In this work, we adopt a more flexible and statistically robust method: \emph{Gaussian Process Regression} (GPR) \citep{mertens2018,Ghosh2020}, which is a non-parametric Bayesian framework for regression. This allows us to model the data as a realization of a Gaussian process characterized by a mean function and a covariance (kernel) function. Unlike the polynomial method, GPR does not assume an explicit parametric form for the foregrounds. Rather, it infers the spectral smoothness directly from the data through the covariance structure. In this framework, the observed visibilities $V(\nu)$ at frequency $\nu$ are treated as a combination of different contributions, each with distinct spectral behaviour. The foreground component is modeled as a smooth function of frequency, with strong correlations extending across large bandwidths, and the 21-cm signal shows coherence only over relatively narrow frequency intervals \citep{Datta-2007a}. This difference in correlation scale allows GPR to disentangle the smooth foregrounds from the fluctuating 21-cm component in a statistically consistent way. It also minimizes the risk of losing the cosmological \HI\ signal.

We model the smooth foreground component as a Gaussian process, given by 
\begin{equation}
    V_{\text{fg}}(\nu) \sim \mathcal{GP}(m(\nu), K(\nu,\nu^\prime)),
\label{eq:gp}
\end{equation}
where $m(\nu)$ is the mean function and $K(\nu, \nu')$ is the covariance function (kernel) that encodes our prior belief about the smoothness and correlation of the signal across frequencies.

 We use the Radial Basis Function (RBF) kernel, also known as the Squared Exponential kernel. The RBF kernel enforces smooth spectral correlations, consistent with the expected behavior of the foreground components. This also prevents the overfitting that occurred in the polynomial fitting method. The covariance between the foreground signal at two frequencies, $\nu$ and $\nu'$, which is chosen as the RBF kernel, can be written as, 
\begin{equation}
    K_{\text{RBF}}(\nu, \nu') = \sigma_f^2 \exp\left(-\frac{(\nu - \nu')^2}{2\ell^2}\right),
    \label{eq:gpr_RBF}
\end{equation}
where $\sigma_f^2$ is the signal variance and $\ell$ is the correlation length scale. The length scale $\ell$ determines how rapidly the correlation between two frequency channels decays as their separation increases. For the foregrounds, which vary smoothly with frequency, $\ell$ is expected to be large, typically corresponding to several MHz.

\begin{figure}
\centering
\includegraphics[width=0.49\textwidth]{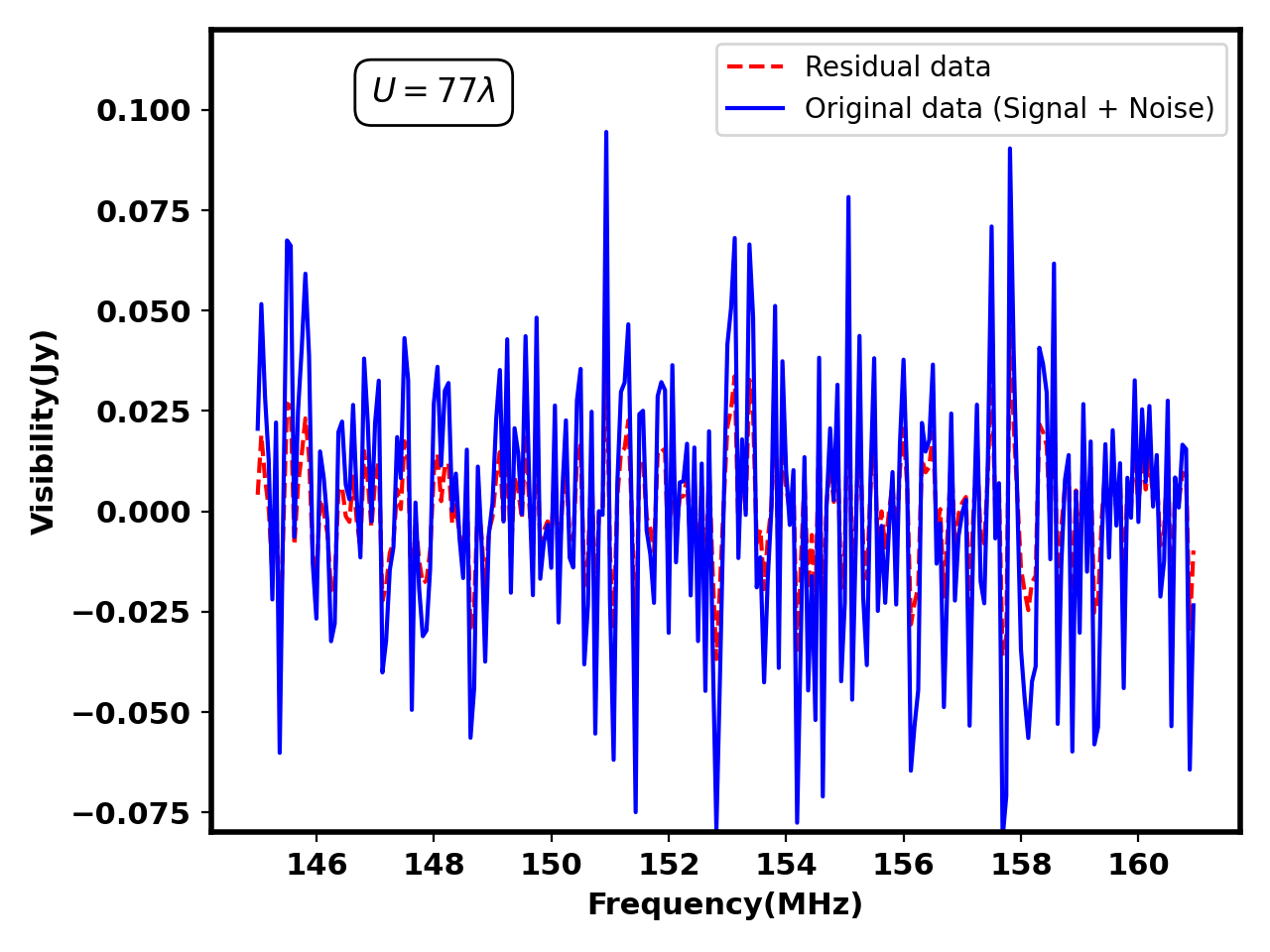}
\caption{This shows the comparison of the recovered visibility (red dashed lines) after subtracting the foreground contribution using the GPR with the original visibility consisting of the HI 21-cm signal and noise contributions (blue lines).}
\label{fig:gpr_vs_poly}
\end{figure}

In our framework, we fix $\ell$ to a large value representative of several MHz rather than optimizing it independently for each baseline.  This choice is motivated by the fact that the dominant foreground components are expected to vary smoothly over a large frequency bandwidth, while the cosmological \HI\ 21-cm signal decorrelates on much smaller frequency scales. Fixing $\ell$ ensures a uniform and stable foreground model across the entire $uv$-plane and avoids baseline-to-baseline fluctuations that can arise from independent hyperparameter optimization.

Figure~\ref{fig:gpr_vs_poly} compares the recovered visibility (red dashed lines) after subtracting the foreground contribution using GPR with the original visibility consisting of the \HI\ 21-cm signal and noise contributions (blue lines). This demonstrates that GPR efficiently removes the smooth foreground component from the mock visibility data without overfitting the \HI\ 21-cm signal.

\section{Matched Filter: Bare estimator}
\label{sec:matched_filter}

Once foregrounds are subtracted, the residual visibility is expected to contain only the faint \HI\ 21-cm signal and the highly dominant system noise contribution from radio interferometers. To enhance the detectability, we employ the matched filtering technique, which was first introduced in \cite{Datta_2007} and later used in our earlier work \cite{Mishra:2024jjg}. In this approach, we construct a filter that follows the functional form of the expected signal. We get maximum signal-to-noise ratio (SNR) when the filter matches the target signal. %This method optimally combines the \HI\ 21-cm signal while suppressing the system noise, thereby enhancing the SNR. 
Since the functional form of the visibility corresponding to a spherical \HII\ region buried in a fully or partially neutral hydrogen medium is known \citep{Datta_2007} and the system noise in each baseline and frequency channel is random and follows a Gaussian distribution, the matched filter framework provides an optimal strategy for detecting ionized bubbles in 21-cm maps. The matched filter estimator can be written as  
\begin{equation}
    {\hat{E}} = \left[{\sum_{a,b} {S_f}^* (\bm{U}_a, \nu_b) \hat{V}(\bm{U}_a, \nu_b)}\right]/\left[{\sum_{a,b}1}\right],
    \label{eq:estim}
\end{equation}
where $S_f(\bm{U},\nu)$ is the matched filter, and $V(\bm{U},\nu)$ is the residual visibility data. The estimator sums the quantity ${S_f}^* (\bm{U}_a, \nu_b) \hat{V}(\bm{U}_a, \nu_b)$ over all the baselines ($U_a$) and frequency channels ($\nu_b$) and averages over them. The system noise contribution to the estimator is expected to be zero when summed over a large number of independent realizations. However, its contribution is unlikely to be exactly zero for a single realization of the system noise. The variance of the estimator can be given by,
\begin{equation}
   \left<(\Delta {\hat{E}})^2\right>_{NS}= {\langle N^2\rangle} \left[\sum_{a,b} \left| S_f(\bm{U}_a, \nu_b) \right|^2\right] / \left[\sum_{a,b} 1\right]^2,
\end{equation}
where ${\langle N^2\rangle}$ is the variance of the system noise for a single visibility, which can be computed from eq. \ref{eq:sys-noise}. The corresponding SNR then can be calculated as,
\begin{equation}
    {\rm SNR}=\frac{\langle \hat{E} \rangle}{\sqrt{ \left<(\Delta {\hat{E}})^2\right>_{\rm NS}}}. 
    \label{eq:SNR}
\end{equation}
We adopt ${\rm SNR} \gtrsim 5$ as the threshold for a robust detection.

\section{Fast estimator: Gridding the data}
\label{subsec: gridding}
We see that, while calculating the estimator using the above method, it is necessary to first subtract the foregrounds from the observed total visibility for each baseline separately. Subsequently, the residual visibilities are multiplied by the filter, and the resulting products are summed over all baselines and frequency channels to compute the estimator. In practice, the SKA1-Low will typically measure visibilities at billions of baselines (for a typical $100$ hours of observation with a $10$ sec integration time) at a single frequency channel. Subtracting the foregrounds for each baseline separately and then estimating the estimator would be highly computationally expensive, requiring substantial computing and human resources. Moreover, constraining the EoR and Cosmic Dawn model parameters using MCMC techniques on such large datasets would be even more computationally demanding. To reduce this computational cost, a more practical approach is to grid the visibilities onto a regular $uv$ plane. It reduces the large data volume and provides a fast estimator for analyzing large datasets. Therefore, we employ the fast estimator which uses the gridded visibilities for different SKA1-Low configurations.

Using the shortest and longest baselines required, we divide the entire u–v plane into a two-dimensional grid. The grid spacing $\Delta u$ is chosen in accordance with the size of the primary field of view ($\theta_0$), following the relation $\Delta u \approx 1 / \theta_0$. In this study $\Delta u = 1 / \theta_{\rm im}$, where $\theta_{\rm im}$ is the angular size (in radian) of the simulated \HI\ 21-cm and foreground map. Here, the maximum baseline is $u_{\rm max}=1/{\theta_{\rm grid}}$, where ${\theta_{\rm grid}}$ is the angular resolution of  simulated image. For simulated \HI\ maps of comoving size of $527$ Mpc and spatial resolution of $1.013$ Mpc, the resulting $u_{\rm max}=4343 \lambda$ and $\Delta u=17\lambda$. %Using this estimator, we do not simulate visibilities directly from the measured SKA1-Low baselines. Instead, we first Fourier transform the model sky brightness temperature maps into visibility space on a regular $u$–$v$ grid, thereby producing a set of gridded visibilities $V_{\rm grid}(u,v,\nu)$. 

To account for the actual SKA1-Low baseline sampling, we simulate baseline distribution for $8$ hrs of observations with $320$ sec integration time using a particular SKA Low array configuration. The raw baselines are then gridded using the \emph{nearest grid point} (NGP) assignment scheme \citep{hockney1988computer,Thompson2017}. In this approach, we map each simulated baseline vector $\mathbf{U}_i=(u_i,v_i)$ at frequency $\nu$ to its nearest grid cell $(u_g,v_g)$. The number of baselines assigned to a given grid point defines the baseline weight,
\begin{equation}
N_g(u_g,v_g) = \sum_{i} W_i(u_g,v_g),
\label{eq:baseline_weight}
\end{equation}
where $W_i=1$ if the $i$-th baseline is mapped to $(u_g,v_g)$, and $W_i=0$ otherwise. Thus, cells densely sampled by the interferometer receive higher weights, while cells with no baselines have $N_g=0$.

We compute the Fourier transforms of the simulated \HI\ 21-cm and foreground maps to obtain their respective contributions to the total visibilities on a gridded baseline plane. %There are three separate contributions in the Fourier-transformed visibilities of the sky maps: the cosmological 21-cm signal ($V_{21\text{cm}}$), the astrophysical foregrounds ($V_{\rm FG}$), and the system noise ($V_{\rm N}$). 
However, the system noise in each baseline is random and expected to follow a Gaussian distribution. To simulate its contribution on each gridded baseline, we generate random numbers drawn from a Gaussian distribution with mean zero and rms given by eq.~\ref{eq:sys-noise}. Each random number is then scaled by a factor $1/\sqrt{N_g}$ to account for averaging over $N_g$ visibilities per grid cell. Consequently, the total visibility at each grid point can be written as,
\begin{multline}
V_{\rm grid}(u_g, v_g, \nu) =
\begin{cases}
V_{21\text{cm}}(u_g, v_g, \nu) + V_{\rm FG}(u_g, v_g, \nu)\\
+\, V_{\rm N}(u_g, v_g, \nu)/\sqrt{N_g(u_g, v_g)}, & N_g > 0, \\[10pt]
0, & N_g = 0.
\end{cases}
\label{eq:visb_gridded}
\end{multline}

% \begin{equation}
% V_{\rm grid}(u_g, v_g, \nu) =
% \begin{cases}
% V_{21\text{cm}}(u_g, v_g, \nu) + V_{\rm FG}(u_g, v_g, \nu) +\\ \dfrac{V_{\rm N}(u_g, v_g, \nu)}{\sqrt{N_g(u_g, v_g)}}, & N_g > 0, \\[10pt]
% 0, & N_g = 0.
% \end{cases}
% \label{eq:visb_gridded}
% \end{equation}
\noindent
This formulation incorporates all contributions consistently to the final gridded visibilities.  

\section{Results: Detectability}
\subsection{Comparison: Bare estimator vs Fast estimator}
\begin{figure}
    \centering
    \includegraphics[width=1\linewidth]{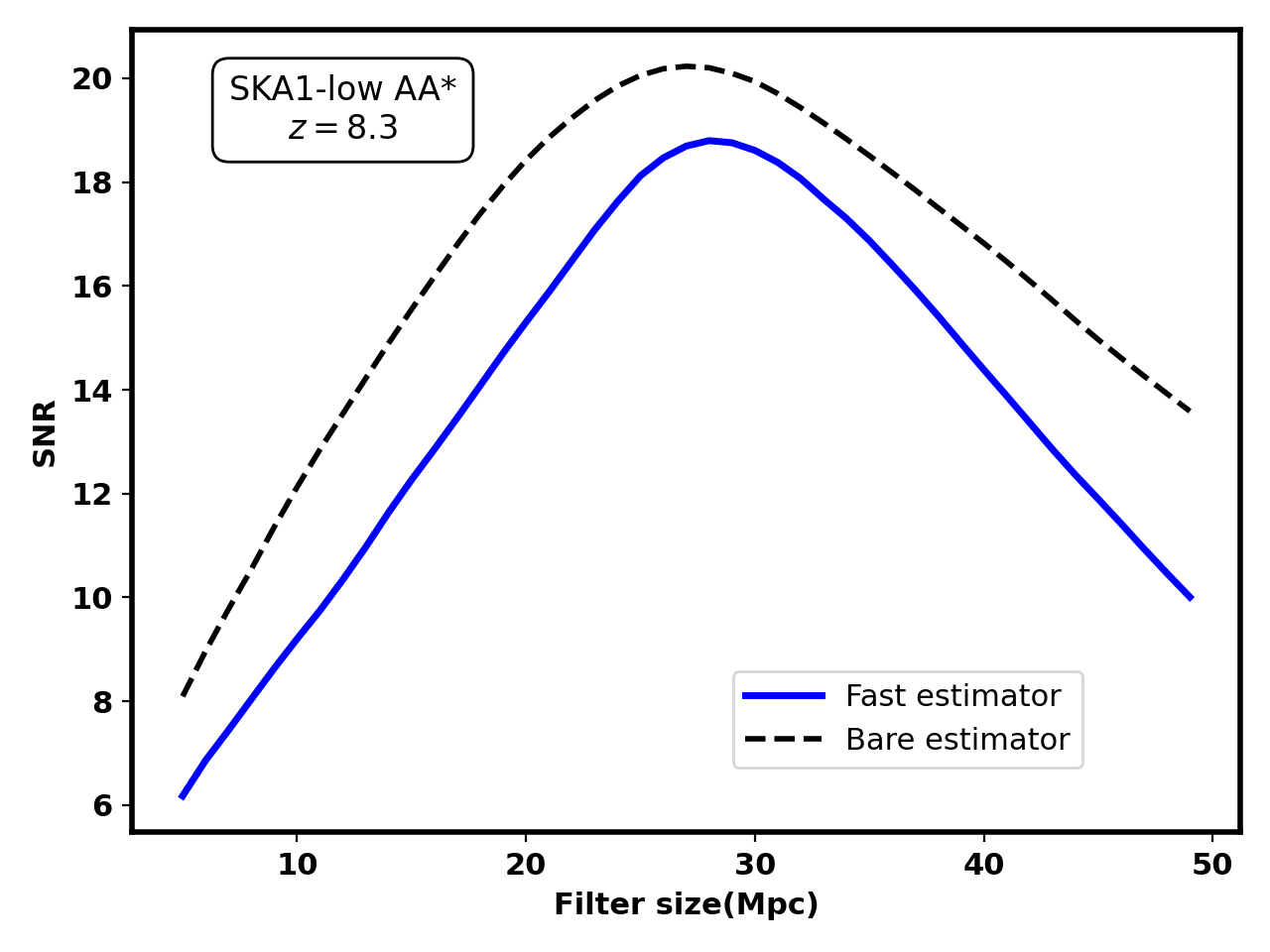}
    \caption{Comparison between the bare estimator(dashed line) and fast estimator(solid line) for the SKA1-Low \AAstar\ configuration at $z=8.3$. The plot shows the signal-to-noise ratio (SNR) as a function of filter size. Both estimators employ GPR-based foreground subtraction method.}
    \label{fig:Bare_vs_Fast}
\end{figure}

Here, we compare the performance and computational efficiency of two different approaches - the bare estimator and the fast gridded estimator - for calculating the estimator and the corresponding signal-to-noise ratio using the matched filter. Both methods utilize the GPR method for foreground subtraction, but differ in how they handle the visibility data. In the case of the bare estimator, the foreground is first subtracted from the total observed visibility for all available baselines. Then, quantities such as $S_f^* \hat{V}$ and $|S_f|^2$ are summed over all baselines and frequency channels to calculate the estimator and its variance for different filter sizes.  In the case of the fast gridded estimator, the visibilities for all baselines are first gridded according to the prescription described in the previous section. This substantially reduces the number of data points and thus the computational cost. The estimator and its variance are then calculated using the gridded data after subtracting the foregrounds. In order to check the accuracy of the fast estimator, we compare the estimated signal-to-noise ratio with that obtained from the more accurate bare estimator.  We consider a scenario with the SKA1-Low \AAstar\ configuration at $z=8.3$ (refer to case II from Subsection~\ref{subsec: HI_sim}). Here, the simulated ionized bubble is aspherical in nature, having an approximate bubble radius of around $28$~Mpc. Figure~\ref{fig:Bare_vs_Fast} shows the SNR as a function of filter size for the bare estimator (dashed line) and the fast estimator (solid line). We get a peak SNR value of $20.3$ at a filter size of $28$~Mpc for the bare estimator. On the other hand, for the gridded estimator, SNR peaks at a filter size of $29$~Mpc with a peak SNR value of $18.5$. We see that the fast gridded estimator reliably predicts both the SNR and the peak location. The slight offsets in these quantities arise mainly from assigning the visibilities to the nearest gridded baselines.

%This slight offset in the peak filter size location arises due to the effective averaging over baselines in the gridded approach, which slightly smooths the signal visibility; consequently, the peak filter shifts to the outer radius of the ionized bubble. However, the overall SNR values obtained from both estimators are comparable, indicating that the gridding procedure does not lead to a significant loss of information.

While the bare estimator yields marginally higher SNR values, it is computationally significantly more expensive. This computational cost scales directly with the number of visibility points processed. For the specific observation setup used in Figure~\ref{fig:Bare_vs_Fast} (SKA1-Low \AAstar with 307 stations), the number of baselines is $N_{b} = 307 \times 306 / 2 = 46,971$.
Over 8 hours of observation with a correlator integration time of $320$~s, the total number of scans is $\approx 90$, resulting in a total visibility count of $\approx 4.2 \times 10^6$ per frequency channel. In contrast, the fast estimator operates on a fixed $512 \times 512$ grid, reducing the effective data points to $\approx 2.6 \times 10^5$. This corresponds to a reduction in computational cost by a factor of $\sim 16$. This disparity becomes even more pronounced for realistic observational scenarios with finer time sampling. For instance, a $10$~s integration time increases the number of scans to $2880$, making the bare estimator nearly $\sim 500$ times more computationally expensive than the fast gridded estimator. This high cost arises from the application of GPR-based foreground modeling for each baseline in order to subtract the foregrounds from visibility data. We therefore adopt the fast estimator for all subsequent analyses, as it provides an optimal balance between accuracy and computational efficiency.

%While the bare estimator yields marginally higher SNR values, it is computationally expensive, requiring nearly $\sim 20$ times {\bf KKD: how do you calculate it?}longer computation time than the fast estimator, making it unsuitable for large SKA datasets. This high cost arises from the application of GPR-based foreground modeling for each baseline in order to subtract the foregrounds from visibility data. We therefore adopt the fast estimator for all subsequent analyses, as it provides an optimal balance between accuracy and computational efficiency.

\subsection{Detectability for SKA1 -Low AA* and AA4}
\label{subsec:SNR_configs}
\begin{figure*}
\centering
\includegraphics[width=0.49\textwidth]{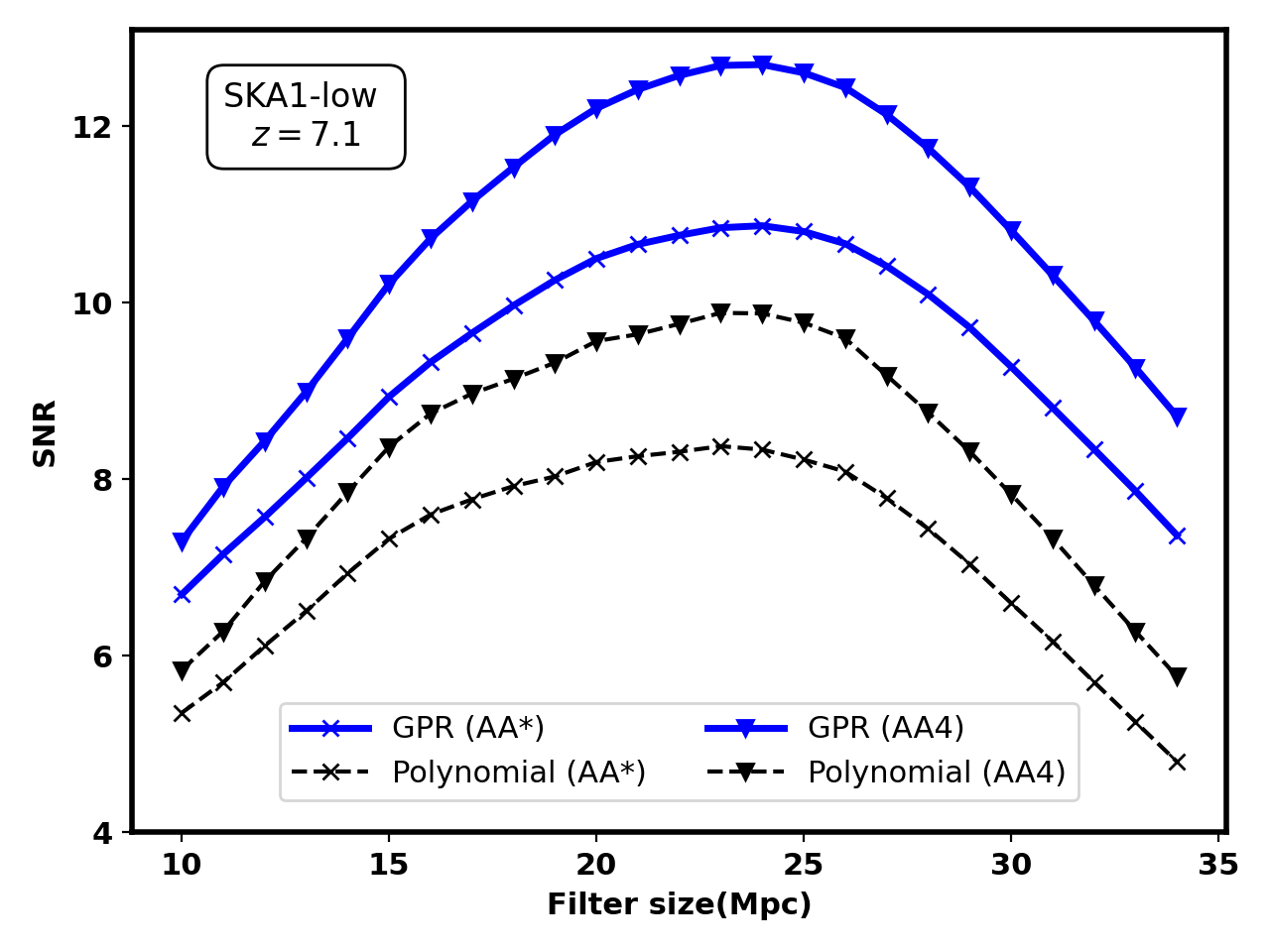}
\includegraphics[width=0.49\textwidth]{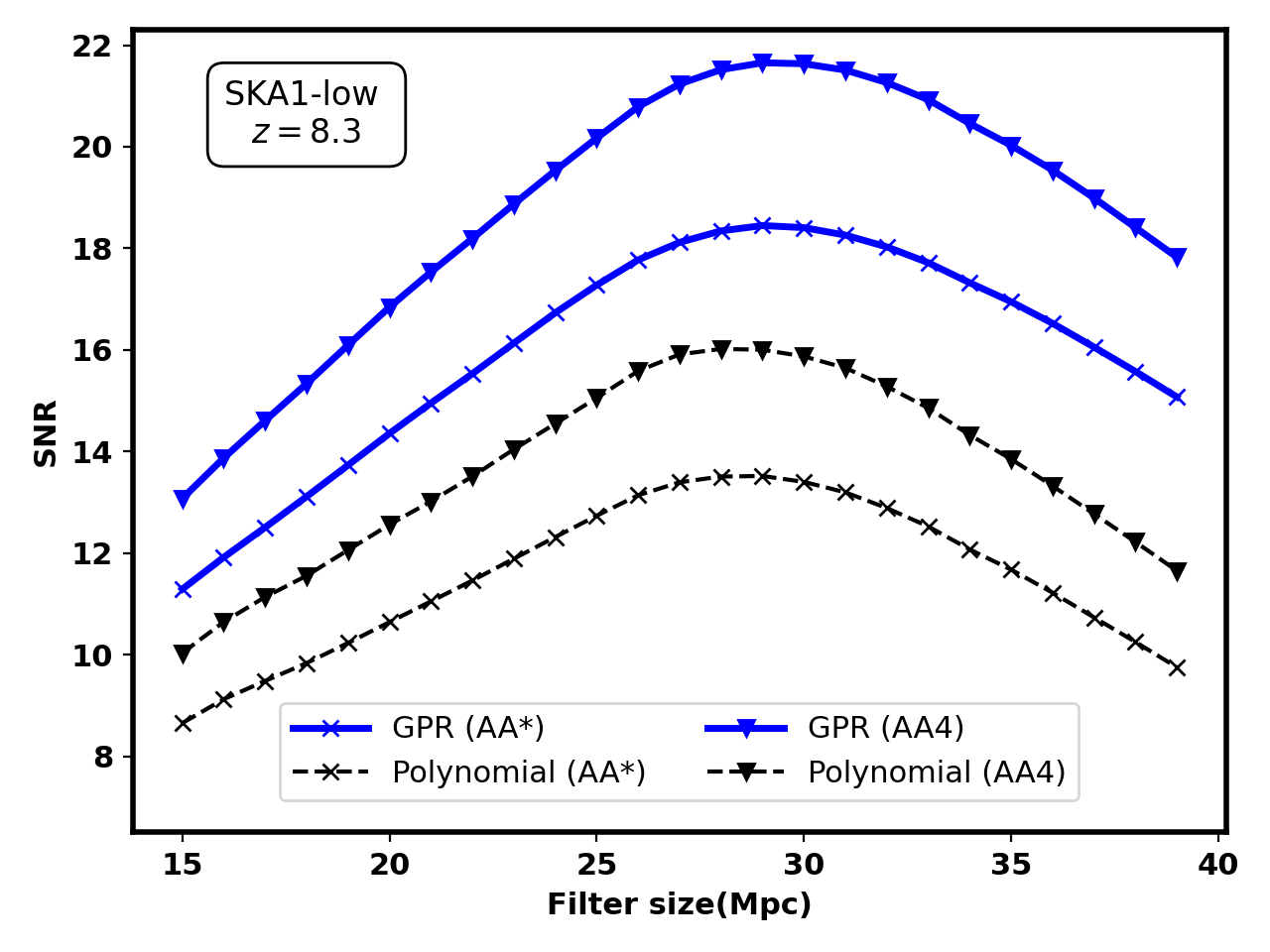}

\caption{SNR as a function of the filter size for ionized bubbles at two different redshifts. The panels compare the performance of GPR (solid lines) and polynomial (dashed lines) foreground subtraction methods for the SKA1-Low \AAstar\ and AA4 configurations.}
\label{fig:SNR_vs_filter_size_gpr}
\end{figure*}

To study the feasibility of detecting individual ionized bubbles using the SKA1-Low AA and AA4 configurations, we compute the signal-to-noise ratio (SNR) using the matched-filter formalism described in Section \ref{sec:matched_filter}. We also examine the impact of foreground subtraction using two different methods, namely GPR and polynomial fitting.

Figure~\ref{fig:SNR_vs_filter_size_gpr} shows the SNR as a function of filter size for two different scenarios as discussed in Subsection~\ref{subsec: HI_sim}. The left panel shows the SNR plots for $100$ hrs of observations for the ionized bubble at $z=7.1$ with outside neutral fraction $x_{\rm HI}=0.88$, and the right panel shows the same at $z=8.3$ with $x_{\rm HI}=0.94$. In each panel, we compare the performance of two foreground subtraction methods, the GPR and polynomial fitting, for the SKA1-Low AA* and AA4 array configurations. From the left panel of Figure~\ref{fig:SNR_vs_filter_size_gpr}, we see that at redshift $z=7.1$ the SNR value peaks at around a filter size of $\sim 23-24$ Mpc for each of the four curves, which is similar to our original input bubble size in the simulation. Similarly, from the right panel of Figure~\ref{fig:SNR_vs_filter_size_gpr}, at redshift $z=8.3$ we get the peak SNR at a filter size of $\sim 28-29$ Mpc, which is close to our original input bubble size in the simulation.

Table~\ref{tab:SNR_results} shows the peak SNR values for different SKA1-Low configurations and for different foreground subtraction methods for the two scenarios. 
\begin{comment}
    \begin{table}[h]
\centering
\caption{Peak SNR values for different SKA1-Low configurations and foreground subtraction techniques at two representative redshifts. Each case corresponds to $100$ hours of total observation time.}
\label{tab:SNR_results}
\begin{tabular}{lcc}
\hline
\hline
Configuration \& Method & $z = 8.3$  & $z = 7.1$ \\
\hline
\AAstar\ (Polynomial fitting)  & $13$ & $8$  \\
\AAstar (GPR subtraction)     & $18$ & $11$ \\
AA4 (Polynomial fitting)  & $16$ & $10$ \\
AA4 (GPR subtraction)     & $22$ & $13$ \\
\hline
\hline
\end{tabular}
\end{table}
\end{comment}
\begin{table}
\centering
\caption{Peak SNR values and corresponding filter sizes for different SKA1-Low configurations and foreground subtraction techniques at two representative redshifts. Each case corresponds to $100$ hours of total observation time.}
\label{tab:SNR_results}
\begin{tabular}{lccc}
\hline
\hline
Configuration \& Method  & Peak SNR & Filter Size (Mpc) & Redshift \\
\hline
\AAstar\ (Polynomial fitting)  & 13.5 & 29 & $z = 8.3$ \\
\AAstar\ (GPR subtraction)     & 18.5 & 29 & $z = 8.3$ \\
AA4 (Polynomial fitting)       & 16 & 28 & $z = 8.3$ \\
AA4 (GPR subtraction)          & 21.7 & 29 & $z = 8.3$ \\
\hline
\AAstar\ (Polynomial fitting)  & 8.4  & 23 & $z = 7.1$ \\
\AAstar\ (GPR subtraction)     & 10.9 & 24 & $z = 7.1$ \\
AA4 (Polynomial fitting)       & 9.9 & 23 & $z = 7.1$ \\
AA4 (GPR subtraction)          & 12.7 & 24 & $z = 7.1$ \\
\hline
\hline
\end{tabular}
\end{table}
In both scenarios, we see that the SNR considerably drops when we apply the polynomial fitting method instead of GPR. This is because when we apply the polynomial fitting method to model the foregrounds, it also models some part of the \HI\ 21-cm signal and consequently subtracts some part of the \HI\ 21-cm signal. This results in a lower SNR peak value. In the case of GPR, we find that it models the smooth foreground component while leaving most of the HI 21-cm signal intact. Consequently, this gives us much higher SNR values. These results confirm our expectation that GPR is more effective at preserving the \HI\ signal while removing foregrounds, as discussed in Section~\ref{sec:FG_subtract}. 

The comparison across array configurations highlights the effect of instrumental sensitivity and baseline coverage. The full AA4 configuration provides the higher SNR due to its larger collecting area and superior $uv$-coverage, which improves sensitivity on both small and large angular scales. The intermediate \AAstar\ configuration, while less sensitive than AA4, still achieves robust detections within $100$ hours of integration. At $z = 8.3$, the maximum SNR reaches $\sim 21.7$ for AA4 and $\sim 18.5$ for \AAstar\ (using GPR), while at $z = 7.1$, the corresponding values are $\sim 12.7$ and $\sim 10.9$, respectively. We see that all cases exceed the nominal $5\sigma$ detection threshold. This shows that both SKA1-Low configurations are capable of detecting ionized bubbles of these sizes within reasonable observation times.

\section{Scaling relation}
\label{sec:scaling_relation}
In the previous sections, we have restricted our detailed analysis to three specific representative scenarios (e.g., a QSO at $z=7.1$ and a galaxy cluster at $z=8.3$). However, the properties of ionized bubbles are largely unknown and may differ significantly from these specific cases. Running full-scale HI 21-cm simulations, foreground subtraction pipeline, matched filter estimation, and parameter estimation pipelines for every possible combination of bubble size, redshift, and neutral fraction is computationally expensive. Therefore, it would be useful to have a general scaling relation that can quickly predict the detectability of ionized bubbles for a wide range of parameters for the SKA1 -Low. This will also help in the planning of future targeted observations.

 A theoretical framework for the scaling of the matched-filter SNR was previously established by \citet{Datta_2009}. It derived an analytical expression for the expectation value of the SNR, which depends on bubble size, neutral Hydrogen fraction, cosmological, and the instrument parameters. The scaling relation can be written as:
\begin{equation}
    \mathrm{SNR} \propto  x_{\rm HI} \frac{A_{\rm eff}}{T_{\rm sys}}\frac{(1+z)}{H(z)}\sqrt{N_{\rm b} t_{\rm obs}\frac{R_{\rm b}^3}{r_\nu^2 r_\nu'}},
    \label{eq:scaling_eq_09}
\end{equation}
where $A_{\rm eff}$ is the effective collecting area, $t_{\rm obs}$ is the observation time, $N_{\rm b}$ is the number of baselines, $T_{\rm sys}$ is the system temperature, $R_{\rm b}$ is the bubble radius, $x_{\rm HI}$ is the neutral hydrogen fraction and the remaining terms describe the background cosmological parameters \citep{Datta_2009}. The scaling relation developed in \citet{Datta_2009} provides a quick method for estimating SNR for the GMRT and the MWA. This scaling relation cannot be directly applied to SKA1-Low because it has a different baseline distribution and a different $A_{\rm eff}(\nu)/T_{\rm sys}(\nu)$. Therefore, we derive separate scaling relations for the two SKA1-Low array configurations, \AAstar and AA4, given their distinct $uv$-coverage. We find that the scaling relation follows the following form:
\begin{equation}
    \mathrm{SNR} \propto \frac{A_{\rm eff}}{T_{\rm sys}} x_{\rm HI} \, (1+z)^{\alpha} \, R_b^{\beta} \, t_{\rm obs}^{1/2}.
    \label{eq:scaling_general_form}
\end{equation}
We find that the SNR is proportional to the square root of the integration time for both array configurations, which is consistent with the theoretical prediction in Eq.~\ref{eq:scaling_eq_09}. Similarly, we find a linear dependence on the mean neutral fraction for both cases. The main differences are seen in the redshift and bubble size scaling, which are influenced by the array configuration and $A_{\rm eff}(\nu)/T_{\rm sys}(\nu)$.  %The fitted scaling relations for SKA1-Low are {\bf KKD: proportional constant?}:
%\begin{itemize}
 %   \item $\alpha = 1.65$ and $\beta = 1.7$ for the \AAstar\ configuration,
  %  \item $\alpha = 1.55$ and $\beta = 1.95$ for the AA4 configuration.
%\end{itemize}
%Combining all the fitted exponents, the final empirical scaling relations for the two SKA1-Low configurations are:
%\begin{align}
%\mathrm{SNR}_{\rm AA^*} &\;\propto\; 
%\frac{A_{\rm eff}}{T_{\rm sys}}x_{\rm HI}\,(1+z)^{1.65}\,R_b^{1.7}\,t_{\rm obs}^{1/2},
%\label{eq:scaling_aastar_final} \\
%\mathrm{SNR}_{\rm AA4} &\;\propto\; 
%\frac{A_{\rm eff}}{T_{\rm sys}}x_{\rm HI}\,(1+z)^{1.55}\,R_b^{1.95}\,t_{\rm obs}^{1/2}.
%\label{eq:scaling_aa4_final}
%\end{align}
The fitted scaling relation for SKA1-Low can be written as,

\begin{equation}
\begin{aligned}
    \mathrm{SNR} =\;
    &K
    \left( \frac{A_{\rm eff}/T_{\rm sys}}{1~\mathrm{m^2 K^{-1}}} \right)
    \left( \frac{x_{\rm HI}}{1} \right)
    \left( \frac{1+z}{10} \right)^{\alpha}
    \left( \frac{R_b}{10~\mathrm{Mpc}} \right)^{\beta} 
    \left( \frac{t_{\rm obs}}{100~\mathrm{hrs}} \right)^{1/2},
\end{aligned}
\label{eq:scaling_SKA}
\end{equation}

where 
\begin{itemize}
    \item $K=3.38$, $\alpha = 1.65$ and $\beta = 1.7$ for the \AAstar\ configuration,
    \item $K=3.10$, $\alpha = 1.55$ and $\beta = 1.95$ for the AA4 configuration.
\end{itemize}

\begin{comment}

\begin{equation}
\begin{aligned}
    \mathrm{SNR}_{\mathrm{AA^*}} =\;
    &K_{\mathrm{AA^*}}
    \left( \frac{A_{\rm eff}/T_{\rm sys}}{1.171} \right)
    \left( \frac{x_{\rm HI}}{0.88} \right)
    \left( \frac{1+z}{8.1} \right)^{1.65} \\
    &\times
    \left( \frac{R_b}{24~\mathrm{Mpc}} \right)^{1.7}
    \left( \frac{t_{\rm obs}}{100~\mathrm{hrs}} \right)^{1/2},
\end{aligned}
\label{eq:scaling_AAstar}
\end{equation}

\begin{equation}
\begin{aligned}
    \mathrm{SNR}_{\mathrm{AA4}} =\;
    &K_{\mathrm{AA4}}
    \left( \frac{A_{\rm eff}/T_{\rm sys}}{1.171} \right)
    \left( \frac{x_{\rm HI}}{0.88} \right)
    \left( \frac{1+z}{8.1} \right)^{1.55} \\
    &\times
    \left( \frac{R_b}{24~\mathrm{Mpc}} \right)^{1.95}
    \left( \frac{t_{\rm obs}}{100~\mathrm{hrs}} \right)^{1/2},
\end{aligned}
\label{eq:scaling_AA4}
\end{equation}
    
\end{comment}

\begin{figure}
\centering
\includegraphics[width=0.5\textwidth]{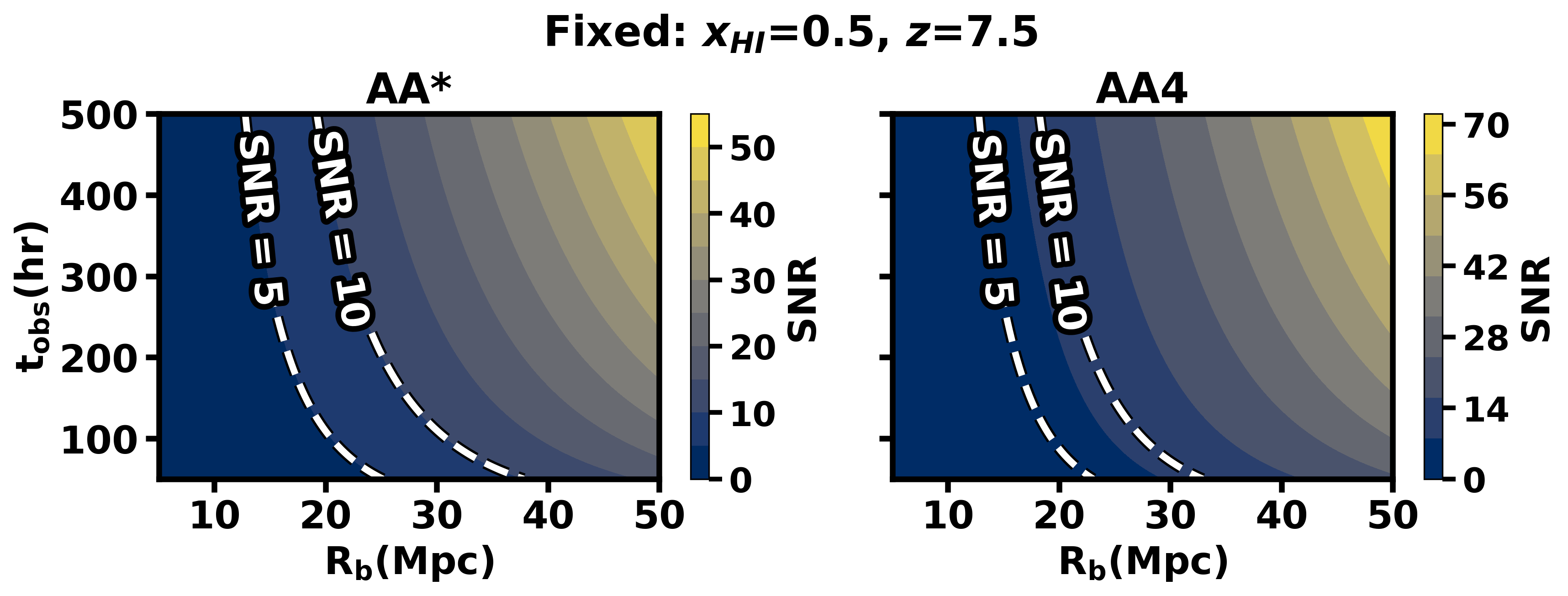}
\caption{The matched-filter SNR as a function of bubble radius ($R_b$) and total observation time ($t_{\rm obs}$) for the $\mathrm{AA}^*$ (left) and $\mathrm{AA4}$ (right) SKA1-Low configurations, fixed at $z=7.5$ and $x_{\rm HI}=0.5$. The dashed white contours indicate the detection thresholds $\mathrm{SNR}=5$ and $\mathrm{SNR}=10$, illustrating the required observation time for detecting bubbles of a given size.}
\label{fig:scaling_tobs_Rb}
\end{figure}

Figure~\ref{fig:scaling_tobs_Rb} presents the predicted signal-to-noise ratio (SNR) as a function of the bubble radius ($R_b$) and the total observation time ($t_{\rm obs}$) for both the $\mathrm{AA}^*$ (left panel) and $\mathrm{AA4}$ (right panel) configurations for $200$ hrs of observations. This analysis is fixed at a redshift of $z=7.5$ and a mean neutral fraction of $x_{\rm HI}=0.5$
. The dashed white contours show the detection thresholds of $\mathrm{SNR}=5$ (robust detection) and $\mathrm{SNR}=10$. 

\begin{figure}
\centering
\includegraphics[width=0.5\textwidth]{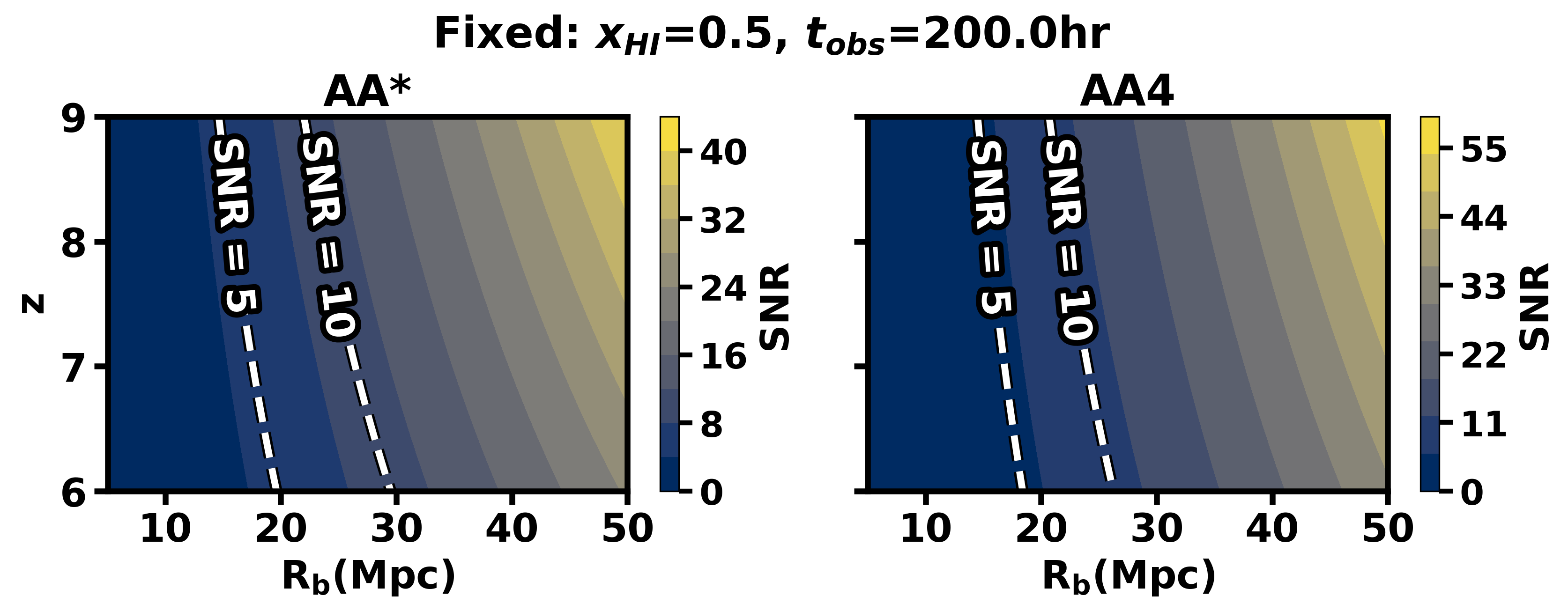}
\caption{The matched-filter SNR as a function of bubble radius ($R_b$) and redshift ($z$) for the $\mathrm{AA}^*$ (left) and $\mathrm{AA4}$ (right) SKA1-Low configurations, fixed at $t_{\rm obs}=200$ hours and $x_{\rm HI}=0.5$. The contours show that the detectability increases strongly with both increasing bubble radius and redshift, with $\mathrm{AA4}$ offering superior performance.}
\label{fig:scaling_Rb_z}
\end{figure}

Figure~\ref{fig:scaling_Rb_z} shows the scaling of the matched-filter SNR across the parameter space spanned by the bubble radius ($R_b$) and the observing redshift ($z$). For this analysis, the observation time is fixed at $t_{\rm obs}=200$ hours and the neutral fraction at $x_{\rm HI}=0.5$. We find that ionized bubbles as small as $\sim 15\,\mathrm{Mpc}$, embedded in a partially ionized medium with $x_{\rm HI}=0.5$, can be detected by both \AAstar\ and AA4 with $200$ hours of observations. Larger ionized bubbles can be detected with significantly shorter integration times.

\begin{comment}
   In an ideal case for uniform antenna separations, we can get the following scaling relation \citep{Datta_2009}:
\begin{equation}
    \mathrm{SNR} \;\propto\; x_{\rm HI}\,(1+z)^{-1/2} \, R_b^{3/2}\, t_{\rm obs}^{1/2}.
\end{equation}

\begin{figure}
\centering
\includegraphics[width=0.49\textwidth]{Images/snr_vs_Rb_AA__AA4.png}
%\includegraphics[width=0.49\textwidth]{Images/snr_vs_z_AA__AA4.png}

\caption{SNR scaling relations.}
\label{fig:scaling_snr}
\end{figure}
\end{comment}

\section{Bayesian analysis}
\label{sec:bayes}
In this section, we assess how precisely the SKA1-Low can constrain the properties of ionized bubbles and the surrounding IGM through a Bayesian analysis. 
We use the likelihood function ($\Lambda$) proposed by \citet{Ghara_2020}. This likelihood function is appropriate for analyses that employ a matched filter to detect a signal buried in strong noise. Our signal is characterized by five parameters: bubble radius ($R$), bubble’s angular coordinates relative to the antenna phase centre ($\theta_x$ and $\theta_y$), bubble’s location along the line of sight, expressed as an offset from the central observing frequency ($\Delta\nu$), and the neutral hydrogen fraction in the surrounding medium ($x_{\rm HI}$). These five parameters can be labeled as, $\mu \equiv [R_f, \theta_x, \theta_y, \Delta\nu, x_{HI}]$. For a uniform prior, the logarithm of the likelihood can be written as \citep{Ghara_2020},
\begin{equation}
     \begin{split}
    \log \Lambda(\mu) = \frac{1}{\sigma_{\rm rms}^2} \int d^2U \int d\nu \, \, \rho_B(\vec{U},\nu)\\
    \quad \left[ 2V(\vec{U},\nu) S_f^*(\vec{U},\nu;\mu)-\left| S_f(\vec{U},\nu;\mu) \right|^2 \right],
     \end{split} 
    \label{eq:log_likelihood}
\end{equation}
where $\rho_B(\bm{U},\nu)$ is the normalized baseline density, defined such that $\int d^2U \int d\nu \, \rho_B(\bm{U},\nu) = 1$, and $\sigma_{\rm rms}$ is the noise rms in image. We explore the parameter space $\mu$ with Markov Chain Monte Carlo (MCMC) methods. The MCMC analysis provides posterior distributions of the parameters, yielding both the best-fit set $\hat{\mu}$ (maximizing the likelihood) and the associated confidence intervals. 

%We also calculate the signal to noise ratio (SNR) in this formalism which  can be written as,
%\begin{equation}
%    \mathrm{SNR} =
%    \left[ \frac{1}{\sigma^2_{\rm rms}} \int d^2U \int d\nu \, \rho_B(\bm{U},\nu) \,\big| S_f(\bm{U},\nu;\mu) \big|^2 \right]^{1/2}.
%    \label{eq:snr_bayesian}
%\end{equation}
%\citep{Finn_1992}. This expression serves as a useful reference for interpreting the detectability of parameter combinations explored in the MCMC chains.

%A point of clarification is required regarding the definition of ``SNR''. In our previous work \cite{Mishra:2024jjg}, the SNR is defined as the ratio of the matched filter estimator ($\widehat{E}$) to its variance ($\Delta\widehat{E}$):
%\begin{equation}
%    \mathrm{SNR}_{\rm MF} = \frac{\widehat{E}}{\sqrt{\Delta\widehat{E}}}. 
%    \label{eq:snr_mf}
%\end{equation}

%In this work we therefore adopt the matched filter SNR (Eq.~\ref{eq:snr_mf}) when discussing detectability and scaling relations, while using the Bayesian detection significance (Eq.~\ref{eq:snr_bayesian}) exclusively for parameter inference.

%\section{Posterior distribution and parameter estimation:}
%\label{subsec:posterior}

We now apply the Bayesian framework 
%detailed in Section~\ref{sec:bayes} 
to the three mock observational scenarios introduced in Subsection~\ref{subsec: HI_sim}. We sample the posterior distributions using the \emph{emcee} \citep{Foreman_Mackey_2013} Markov Chain Monte Carlo (MCMC) sampler using $50$ walkers, each with $50000$ steps. We present the posterior distributions for the five parameters $\mu \equiv [R_{f}, \theta_{x}, \theta_{y}, \Delta\nu, x_{\rm HI}]$ for each case, obtained using the SKA1-Low \AAstar and AA4 configurations.

\begin{figure*}
\centering
\includegraphics[width=0.7\textwidth]{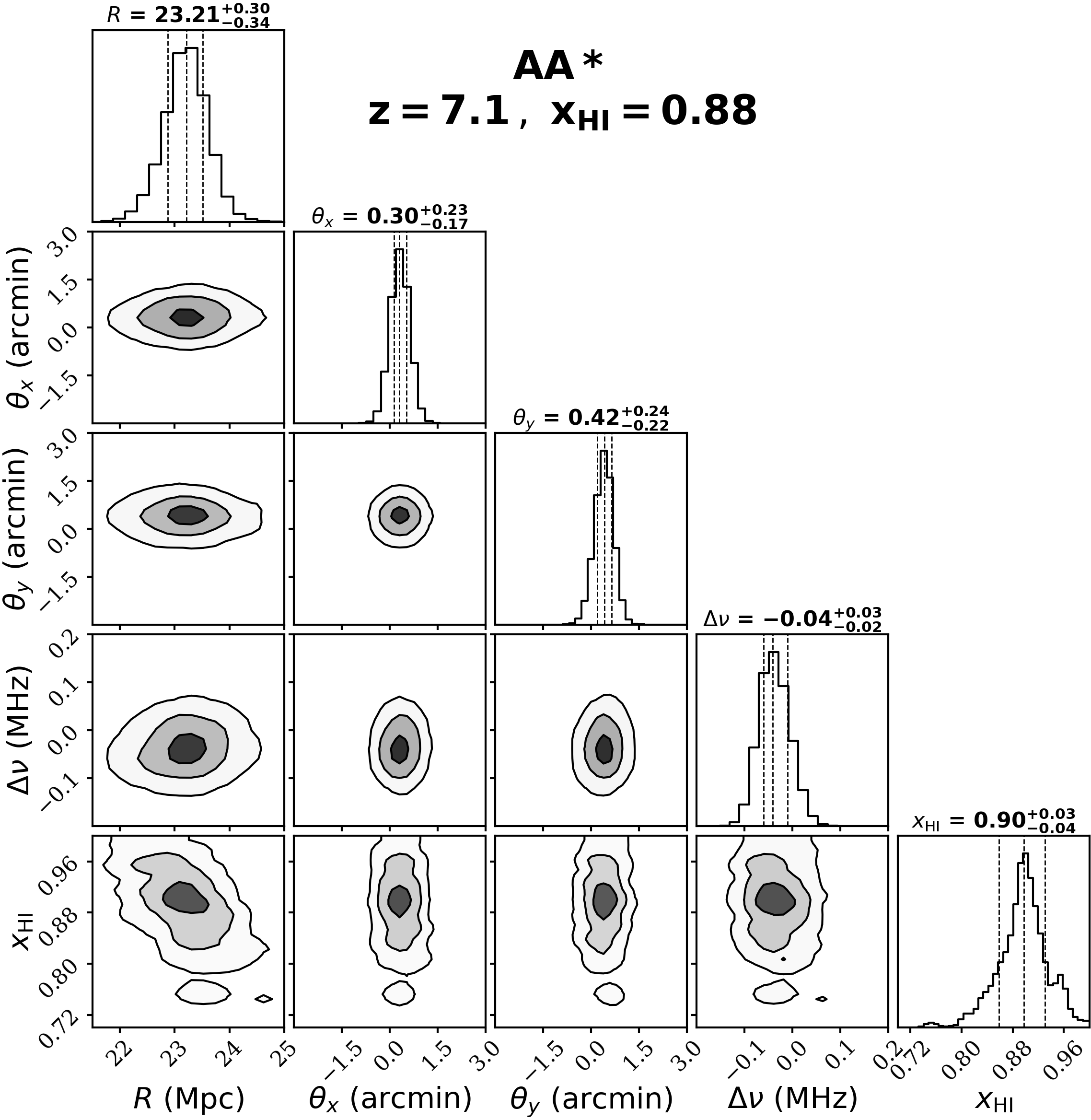}
\caption{Posterior probability distributions of ionized bubble parameters derived from Bayesian analysis of the matched-filter outputs for SKA1-Low \AAstar\ at redshift $z=7.1$. The contours represent the $1\sigma$, $2\sigma$, and $3\sigma$ credible intervals, constraining bubble size, angular and line of sight positions, and neutral hydrogen fraction.}
\label{corner_71_88}
\end{figure*}

Figure~\ref{corner_71_88} shows the posterior distribution of the first scenario at redshift $z=7.1$ with $x_{\rm HI}=0.88$ using the SKA1-Low \AAstar. The recovered bubble radius and associated uncertainty is  $R_{f} = 23.21^{+0.30}_{-0.34}$~Mpc. This is in excellent agreement with the input bubble size of $\sim24$~Mpc. 
The neutral fraction is constrained to $x_{\rm HI} = 0.90^{+0.03}_{-0.04}$, consistent with the input value of $0.88$. The bubble's angular locations ($\theta_{x}, \theta_{y}$) and frequency offset ($\Delta\nu$) are also precisely recovered, with their posteriors centered on their true values.

\begin{figure*}
\centering
\includegraphics[width=0.7\textwidth]{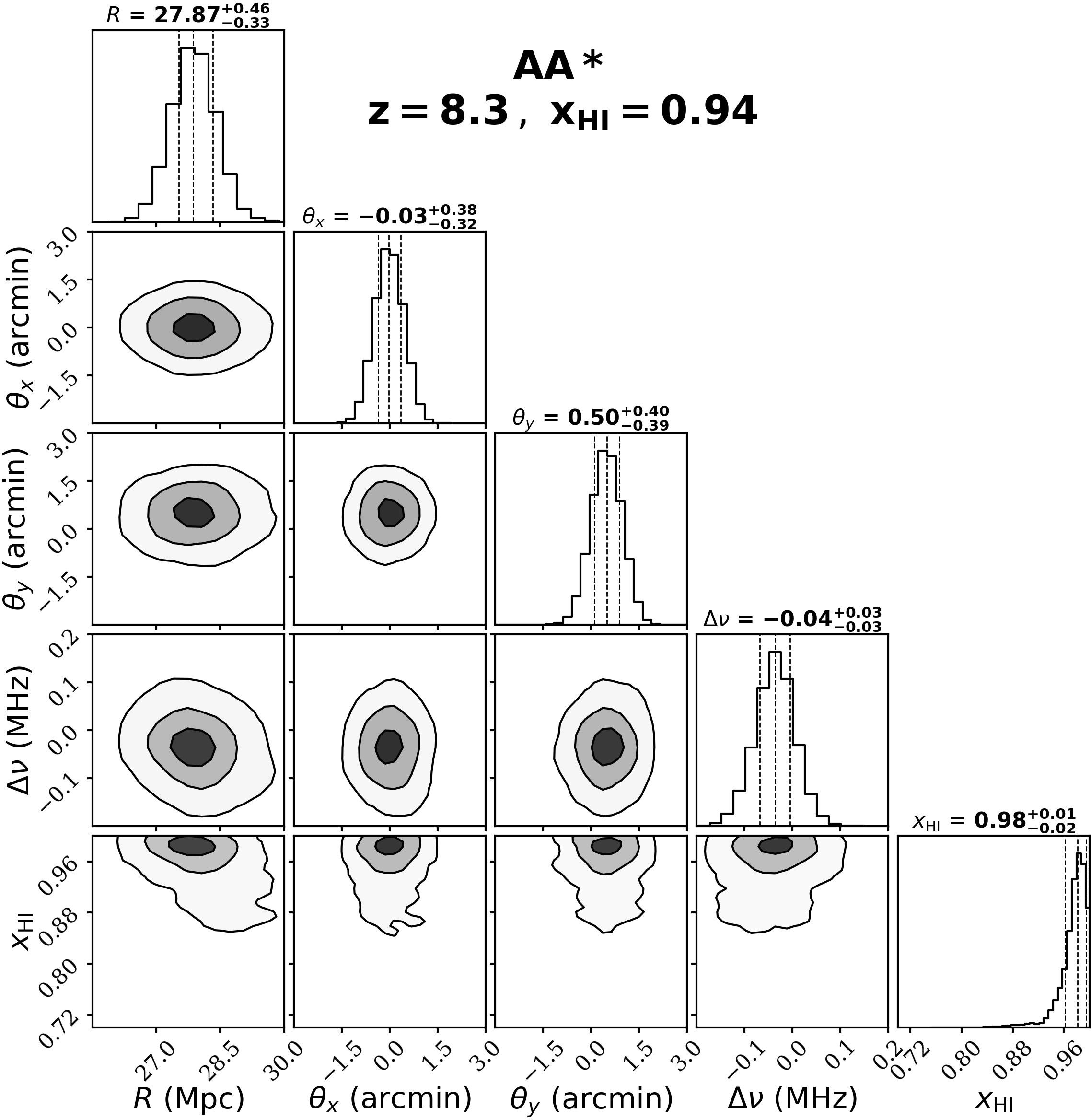}
\caption{Posterior probability distributions of ionized bubble parameters derived from Bayesian analysis of the matched-filter outputs for SKA1-Low \AAstar\ at redshift $z=8.3$. The contours represent the $1\sigma$, $2\sigma$, and $3\sigma$ credible intervals, constraining bubble size, position, frequency offset, and neutral hydrogen fraction.}
\label{corner_83_AAstar}
\end{figure*}

\begin{figure*}
\centering
\includegraphics[width=0.7\textwidth]{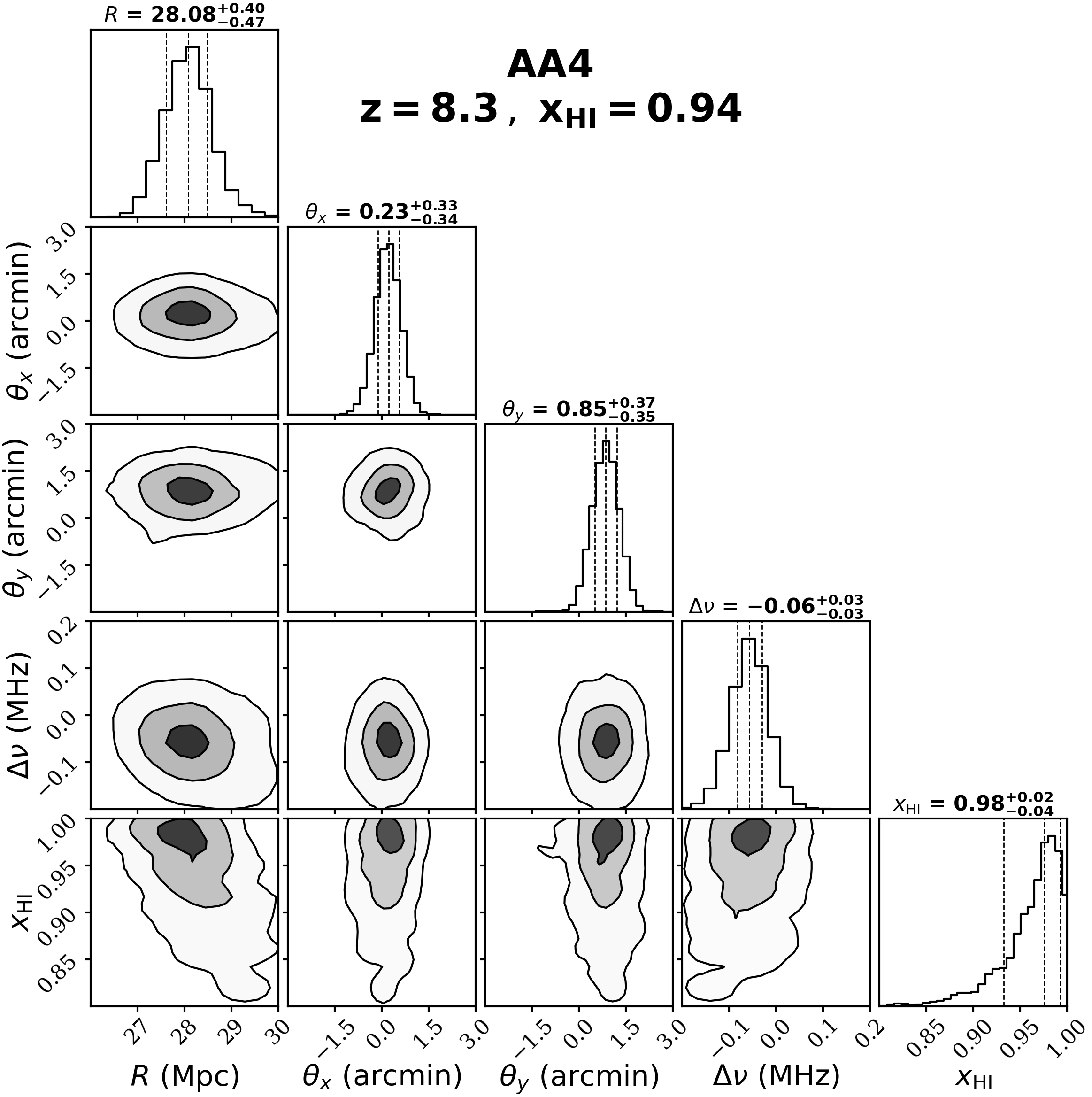}
\caption{Posterior probability distributions of ionized bubble parameters derived from Bayesian analysis of the matched-filter outputs for SKA1-Low AA4 at redshift $z=8.3$. The contours represent the $1\sigma$, $2\sigma$, and $3\sigma$ credible intervals, constraining bubble size, position, frequency offset, and neutral hydrogen fraction.}
\label{corner_83_AA4}
\end{figure*}

Figure~\ref{corner_83_AAstar} $\&$ \ref{corner_83_AA4} show posterior results for the second scenario ($z=8.3$, input $x_{\rm HI}=0.94$) respectively for \AAstar and AA4. For the \AAstar configuration, the recovered bubble radius is $R_{f} = 27.87^{+0.46}_{-0.33}$~Mpc and the neutral fraction is $x_{\rm HI} = 0.98^{+0.01}_{-0.02}$. In comparison, the full AA4 configuration yields $R_{f} = 28.08^{+0.40}_{-0.47}$~Mpc and $x_{\rm HI} = 0.98^{+0.02}_{-0.04}$. Both configurations recover the input radius of $\sim28$~Mpc with high precision. The posterior distributions for both arrays are narrow and unimodal, confirming a robust recovery of all five parameters. 

%The posterior distributions are narrow, confirming a robust recovery {\bf KKD: not clear??. is AA4 better? quote full results including uncertainties}. The recovered radius is consistent with the input size of $\sim28$~Mpc for both the SKA1-Low arrays. The posterior distributions also show tight constraints for the other three parameters.

\begin{figure*}
\centering
\includegraphics[width=0.7\textwidth]{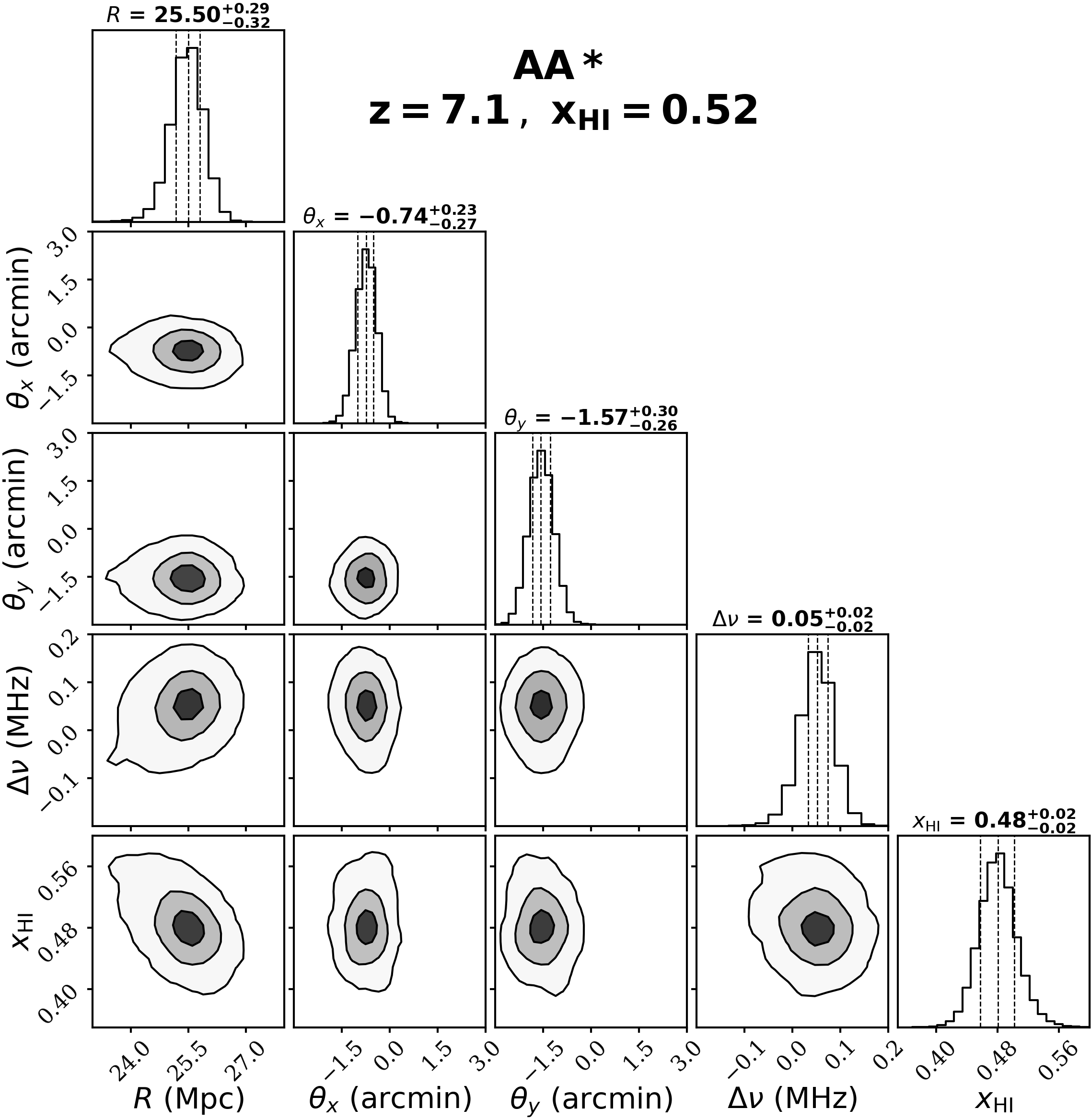}
\caption{Posterior probability distributions of ionized bubble parameters derived from Bayesian analysis of the matched-filter outputs for SKA1-Low \AAstar\ at redshift $z=7.1$ and \XHI$=0.52$. The contours represent the $1\sigma$, $2\sigma$, and $3\sigma$ credible intervals, constraining bubble size, position, frequency offset, and neutral hydrogen fraction.}
\label{corner_71_52}
\end{figure*}

In the more complex environment with $x_{\rm HI}=0.52$ at $z=7.1$ (see the bottom panel of Figure 1 from \citet{Mishra:2024jjg}), we see that the posterior distribution (refer to Figure~\ref{corner_71_52}) shows excellent constraints on all the parameters. Overall, the posteriors remain unimodal and well-behaved across all scenarios. The bubble radius and neutral fraction are reliably constrained in every case, and the angular and line of sight positions are reliably recovered. These results demonstrate that SKA1-Low \AAstar\ and AA4 can not only detect individual ionized bubbles around known luminous sources but also accurately constrain bubble and IGM parameters with $\sim 100$ hours of observations.

\section{Summary \& Discussion}
\label{sec:summary}
Several luminous QSOs and galaxies have now been discovered in recent times at $z \gtrsim 6$. This implies the presence of large ionized regions around them during the epoch of reionization. Detecting and characterizing these individual ionized bubbles using the redshifted \HI\ 21-cm signal offers a direct probe of the reionization epoch. This also complements the widely explored approach of studying the epoch through statistical measures such as the power spectrum \citep{mellema2006, mesinger2016}  or bispectrum \citep{majumdar2018, nasreen25}. In this work, we have developed a fast estimator, employed a non-parametric foreground-subtraction method based on Gaussian process regression, and carried out a detailed investigation of the detectability and parameter estimation of ionized bubbles using SKA1-Low \AAstar\ and AA4 observations. As a demonstration, we have focused on two specific cases: a very bright QSO at $z=7.1$  \citep{Mortlock_2011}, and a large ionized bubble associated with a collection of galaxies at $z=8.3$ \citep{witstok24}. Finally, we have derived a scaling relation that enables rapid estimation of detectability for ionized bubbles of arbitrary size and neutral hydrogen fraction at any redshift with SKA1-Low configurations. 

We have first simulated the \HI\ 21-cm signal around bright reionizing sources and realistic astrophysical foreground components within a 3D cube. We have then generated the complex visibilities using the baseline distributions of the SKA1-Low \AAstar and AA4 configurations, and added system noise contributions based on the telescope sensitivity at the relevant frequencies.  Subsequently, we have applied the GPR method to subtract the foreground contributions from the mock visibilities, which is a significantly more effective method for foreground subtraction than the polynomial fitting method that we have used in our previous work. The polynomial method was found to overfit the data, leading to partial subtraction of the \HI\ 21-cm signal itself and a substantial loss of signal-to-noise ratio (SNR). As shown in Table~\ref{tab:SNR_results}, the peak SNR for the SKA1-Low AA4 configuration at $z=8.3$ dropped from $\sim22$ to $\sim16$ when using polynomial fitting instead of GPR. Results are similar for other cases considered. This shows that GPR successfully models the spectral smoothness of the foregrounds without removing the cosmological \HI\ signal.

After subtracting the foreground contributions from the mock visibility data, we have applied the visibility-based matched filtering method on the residual visibility to detect the ionized bubbles. In this method, the filter optimally combines the \HI\ 21-cm signal around individual ionized regions, maximizing the signal-to-noise ratio. We have further presented a fast estimator for the matched filter, which acts on gridded visibilities, unlike the bare estimator we have used in our previous study. While the bare estimator yields a marginally higher SNR (e.g., $\sim$20.3 vs. $\sim$18.5 for the $z=8.3$ case), it is more computationally expensive. The fast gridded estimator provides an excellent balance between accuracy and computational efficiency. This makes it the practical and necessary choice for analyzing the large datasets expected from the SKA1 -Low. Combining the GPR method and the fast estimator, we studied the detectability of the ionized bubbles for two different scenarios and found that ionized bubbles at $z \approx 7$–$8$ can be detected with SNR $\gtrsim 10$ in $\sim100$ hours of SKA1-Low \AAstar/AA4 observations. We have also presented how the matched–filter SNR scales with the bubble radius, redshift, total observing time, and the mean neutral fraction of the surrounding IGM. This relation can be used to predict the required observation time for different scenarios for SKA1-Low observations.

%we compared them for different SKA1-Low array configurations. The number of short baselines in the SKA1-Low \AAstar and AA4 configurations is very similar; however, $AA4$ includes significantly more long baselines. Consequently, we observed that the difference in the SNR between the two configurations is small, since the HI 21-cm signal contributes very little at large baselines. However, the long baselines play an important role in foreground mitigation, particularly for modelling and subtracting point sources. 

Finally, we have performed Bayesian parameter estimation to recover ionized bubble parameters directly from the residual visibility data. Using an MCMC-based parameter estimation technique, we successfully recovered the bubble radius, position, and mean neutral fraction for representative scenarios at $z = 7.1$ and $z = 8.3$. The posterior distributions consistently peak near the input parameters. This demonstrates that accurate parameter inference is feasible for SKA1-Low \AAstar and AA4 observations with only $\sim 100$ hours of observation time.

While our results demonstrate the potential for bubble detection and characterization with SKA1-Low, we note some limitations in the current study. Our mock observations assume ideal calibration and do not incorporate complex instrumental systematics such as beam chromaticity, direction-dependent effects, or residual RFI, all of which may complicate foreground subtraction in real data \citep{saikat2025}. Furthermore, our matched-filter and Bayesian framework primarily utilize a nearly spherical bubble template. Although we have shown that bubble radii are reliably recovered for different scenarios, including a complex patchy environment, highly irregular or aspherical bubble detection could introduce biases in bubble recovery. Future studies will focus on incorporating realistic instrumental errors and more diverse reionization topologies to further validate our ionized bubble detection pipeline.

\section*{Acknowledgements}
AM  acknowledges financial support from Council of Scientific and Industrial Research (CSIR) via  CSIR-SRF fellowships under grant no. 09/0096(13611)/2022-EMR-I. KKD acknowledges financial support from ANRF
(Govt. of India) under the ARF program (File Number: ANRF/ARG/2025/004594/PS). CSM acknowledges financial support from the Council of Scientific and Industrial Research (CSIR) via a CSIR-SRF Fellowship (Grant No. 09/1022(0080)/2019-EMR-I) and from the ARCO Prize Fellowship. IN acknowledges financial support from the DST-WISE Fellowship (DST/WISE-PhD/PM/2023/104), Govt. of India. SS acknowledges financial support by the Junior Research Fellowship of the University Grants Comission, Govt. of India under Ref. No. 231620034527. The computation work is performed using the facility procured through the financial support of DST-FIST program, Govt. of India provided to the Department of Physics, Jadavpur University vide sanction no. SR/FST/PS-1/2022/219(C).

%%%%%%%%%%%%%%%%%%%%%%%%%%%%%%%%%%%%%%%%%%%%%%%%%%
\section*{Data Availability}
The data underlying this work will be shared upon reasonable re-
quest to the corresponding author.

%%%%%%%%%%%%%%%%%%%% REFERENCES %%%%%%%%%%%%%%%%%%

% The best way to enter references is to use BibTeX:

\bibliographystyle{mnras}
\bibliography{biblio} % if your bibtex file is called example.bib

% Alternatively you could enter them by hand, like this:
% This method is tedious and prone to error if you have lots of references
%\begin{thebibliography}{99}
%\bibitem[\protect\citeauthoryear{Author}{2012}]{Author2012}
%Author A.~N., 2013, Journal of Improbable Astronomy, 1, 1
%\bibitem[\protect\citeauthoryear{Others}{2013}]{Others2013}
%Others S., 2012, Journal of Interesting Stuff, 17, 198
%\end{thebibliography}

%%%%%%%%%%%%%%%%%%%%%%%%%%%%%%%%%%%%%%%%%%%%%%%%%%

%%%%%%%%%%%%%%%%% APPENDICES %%%%%%%%%%%%%%%%%%%%%

% Don't change these lines
\bsp	% typesetting comment
\label{lastpage}
\end{document}